\documentclass[12pt]{amsart}

\usepackage{amsmath}
\usepackage{amsfonts}
\usepackage{amssymb}
\usepackage{amsthm}

\DeclareMathOperator{\alg}{alg}

\DeclareMathOperator{\salg}{step-alg}
\DeclareMathOperator{\selfed}{self-ed}
\DeclareMathOperator{\imsuc}{im-suc}
\DeclareMathOperator{\hist}{hist}
\DeclareMathOperator{\last}{last}

\title{Recursion, Evolution and Conscious Self}
\author{A. D. Arvanitakis}
\email{aarva@math.ntua.gr}
\address{National Technical University of Athens\\ Department of Mathematics\\ Athens, Greece}

\begin{document}
	
\begin{abstract}
	We introduce and study a {\em learning theory} which is roughly {\em automatic}, that is, it does not require but a minimum of initial programming, and is based on the potential computational phenomenon of {\em self-reference}, (i.e. the potential ability of an algorithm to have its program as an input). 
	
	 The conclusions agree with scientific findings in both biology and neuroscience and provide a plethora of explanations both (in conjunction with Darwinism) about {\em evolution}, as well as for the functionality and learning capabilities of {\em human brain}, (most importantly), as we perceive them in ourselves. 
\end{abstract}
	
\maketitle

\section{Introduction}
Before we go into details about the potential use of {\em self-reference}, we give a short argument which favors it:  

We call an executable program $c$ {\em self-editing} if its algorithm $C$ has $c$ as an input and outputs one (or more as we will see) executable programs. Assume now that we have a tree of self-editing executable programs upon which some form of selection (either artificial or natural) applies. Thus the algorithms of every self-editing executable program in the tree is in charge of two activities: To answer to the environment and (as self-editing) to compute their descendants having as an input their executable program. Assuming now Darwinism, the answers to the environment should evolve, thus the system can learn how to answer to the environment. On the other hand, since algorithms produce also their descendants, due to selection, they should learn how to calculate them so as to answer better to the environment. This way the self-editing property of this system serves as a means not only to {\em learn}, but also to {\em learn how to learn,} resulting thus in a much more rapid evolution.

In the classical context of mathematics, {\em self-reference} is a key ingredient of many famous proofs that arrive at a contradiction. One can sketch an outline of this method very briefly, as the application of a function to an argument which is essentially a description of the function itself. The original inventor of the method is a well known mathematician, namely {\em G. Cantor} in his well known proof about the existence of different cardinalities of the infinite \cite{moschovakis}. The method has since appeared also in a lot of equally famous proofs. Among them the {\em incompleteness theorem} of G\"odel and the {\em Halting problem} of Turing. A detailed study on the subject is provided by R.M. Smullyan in \cite{smullyan}.

Especially Turing's approach in the Halting problem, concerns a {\em computable} function (i.e. roughly a function that can be calculated by a personal computer), having as an input its own program.

Henceforth, we are going to use the general term {\em code} to refer either to an executable program, or data, or a unit of both.

Our approach in this paper is slightly different from the above mentioned classic method both regarding the outcome, (as we 'll see later we use this method constructively and not towards the purpose of arriving at a contradiction), as well as in the method itself. We specifically consider computational steps in such a procedure: Given an executable program $c$ for an algorithm, a {\em computational step} of $c$ is determined by an {\em active instruction} of $c.$ To explain things better, assume that the algorithm $\alg (c)$ that $c$ codes for, is given an input $i$ and starting from this input, it computes successively
\[ i = i_1, i_2, \dots, i_n,\]
giving $i_n$ as an output. A {\em computational step} of $\alg(c)$ is any intermediate computation $i_k \mapsto i_{k+1},$ $k< n,$ in which an instruction of $c$ is activated. Such an instruction is generally considered simpler from the algorithm itself, in this case it could be for example to add 1 to the code calculated so far (i.e. to $i_k,$ and thus obtaining $i_{k+1} = i_k + 1.$). We are not going to deal here with what is and isn't {\em simple} enough to count for an instruction, as we are going to use this concept relatively to the corresponding problem at hand. The situation is well known in computer programming, in which instructions may vary from basic ones of the programming language (or the assembly language for that matter), all the way to complex functions already defined to fit our purposes.

Since we deal with activated instructions, (which are subcodes of the original code), it is very convenient to be able to represent the relation of being a subcode to our notation. Thus, we write $c = c[b]$ in order to denote that $b$ is a subcode of $c.$ Notice that in this case $b$ is actually contained in $c.$ For example the code $b  = [0,1]$ is a subcode of $c=[[0,1], 0],$ so that we may write $c[b]$ instead of $c$ to designate that. What we gain this way, is that we may now denote the activated instruction. For example, if $b$ is the activated instruction of $c,$ then we can use a bar above $b$ to demonstrate that. We write in this case that $c$ is in the {\em state} $c[\overline{b}].$

If $b$ is a subcode of $c,$ that is $c = c[b],$ the resulting code if we subtract $b$ from $c,$ will be denoted as $c[\varnothing].$

Using these notations, the subject of our study is sequences of executable programs
\[ c_1, c_2, \dots, c_n,\]
where for every computational step $c_k \mapsto c_{k+1},$ $k < n,$ $c_k$ serves both as the (state-)program  of the step, and also as an input.

In order to distinguish these kind of transitions from the ones found in classical contexts of mathematics, we will be using the term {\em self-editing transitions}. The executable programs $c_k,$ $k< n$ will be called also {\em self-editing} codes. The act of a self-editing transition of a (state-) program $c,$ will be denoted by $\selfed(c).$ Thus, if $c \mapsto c'$ is a self-editing transition, and $c$ is in the state $c[\overline{b}],$ then $c'$ results as the computation of the algorithm $\alg(b),$ that $b$ codes for, with input $c[\overline{b}],$ that is, 
\[ c' = \alg(b)(c[\overline{b}]) = \selfed(c[\overline{b}]).\]

Since $b$ as an instruction can be essentially arbitrary, we obtain the main tool of our study\footnote{We have used here {\em Church-Turing thesis}, which is another way to say that any algorithm has an executable program}:

\smallskip
{\bf Basic Self-Editing Principle.} {\em For every algorithm $B,$ there is an executable program $b,$ (which is in fact a code for $B$) such that for every code $c[\varnothing],$
	\[ \selfed(c[\overline{b}]) = B(c[\overline{b}]).\]}

In order to grasp the meaning of the above principle, it will be helpful to freely translate it as follows:

{\em Assume we are determined to make some changes in an executable program $c[\overline{b}].$ As long as these changes can be expressed in the form of an algorithm (that is $B$), the same changes can be done by $c[\overline{b}]$ to itself, assuming that its computational step is a self-editing one and $b$ is a code for $B.$} 

\smallskip
To see this principle in action, notice that:

\smallskip
{\bf Essential capabilities of a self-editing code:}

	{\bf (1)} {\em A self-editing code may proliferate itself during a self-editing step.} This results from the above principle, by choosing $B$ to output multiple codes. (Subsection \ref{selfReplication}, page \pageref{selfReplication}).
	
	\smallskip
	{\bf (2)} {\em A self-editing code may potentially organize or change its structure.} Suffice it to choose in the above principle an appropriate algorithm $B$ that performs the corresponding changes in the structure of its input. (See for example  subsection \ref{storingDecisions} in page \pageref{storingDecisions})
	
	\smallskip
	{\bf (3)} {\em A self-editing code may retain the memory of itself.} To roughly see this in an one-step calculation, assume that $A$ is any algorithm and define $B(A)$ to be an algorithm that outputs in a combined structure the output of $A$ together with the input. If for example $c' = A(c),$ then $B(A)$ may output the structure $[c, c']$ that contains both $c'$ and $c.$ The result follows by the above Self-editing principle, using $B(A)$ in the place of $B.$ (Section \ref{selfMemory}, page \pageref{selfMemory}).
	
	\smallskip
	{\bf (4)} {\em A self-editing code may add an instruction to itself and moreover it can do it, so that this instruction acts hereditarily.} The capability to add such an instruction follows from the fact that the algorithm $B$ in the above principe can do so. To do it in such a way that it acts hereditarily is somehow more complicated and although several examples that follow in the next sections show that it is possible, the full statement will be addressed in the subsection \ref{permanentDecisions}, page \pageref{permanentDecisions}.
	
	\smallskip
	{\bf (5)} {\em An activated instruction $\overline{b}$ of a self-editing code may alter parameters or subcodes of itself.} It suffices to choose $B$ to act on parameters or components of the subcode $\overline{b}$ of $c[\overline{b}]$. (See also the 2nd example in page \pageref{selfAction}. We are going to use this feature essentially after page \pageref{diagonalization}.)

\smallskip
Granting the above potential capabilities, let us now try to outline our learning theory: 

First notice that due to the capability (1), a self-editing code, may produce more than one such codes, the same can happen to each one of them, and so on. Continuing this procedure we may consider the actual outcome of a self-editing code to be rather a tree than a sequence of self-editing codes. Consequently, we may assume some kind of {\em selection}, either artificial or natural and talk about surviving branches or sequences. Assume now we observe such a surviving sequence
\begin{equation} \label{survivingSequence} c_1, c_2, \dots, c_n.\end{equation}
(Notice at this point, that due to capability (3) of retaining the memory of itself, we may assume that the algorithm of the state code $c_n,$ has as an input the entire sequence \eqref{survivingSequence}, exactly as we are able to observe it).
It is possible then to come to conclusions about what is preferable by the artificial or natural environment, by just noticing details about the structure of each code of the sequence. Assuming as a naive example that in some part of the structure of each $c_k,$ $k \le n,$ the number $k$ is stored, we may conclude that all the descendants of $c_n$ should register $n+1$ in the corresponding part, even if we don't really know anything about the environment. The actual cause of this preference of the environment, may be two-folded: Either we are actually talking about an environment that changes smoothly its preference and to comply with this change we have to add 1 to the aforementioned part of the structure, or increasing the number that is stored there improves the fitness of our codes. In either case, the suggested strategy would be to comply with this, so that if we wanted to enhance the  survivability of our codes, we should insert in $c_n$ an hereditary instruction to augment by 1 this integer in the particular part of the codes of each descendant.

Another naive but essential example, is that we can notice that specific parts of the codes $c_1, \dots, c_n$ remain the same, although this was not intended from the beginning and therefore was not true for every branch of the tree. The fact that we are talking about a surviving branch, may give us in this case the indication, that we should reprogram $c_n$ adding a hereditary instruction to keep intact these parts in every possible descendant. The reason is that this seems to be preferable by the environment, which seems to judge any alteration of the specific subcodes as non-surviving.

Assuming now that we are working on a computer, we can automate this procedure by using some algorithm that performs some kind of pattern recognition, in order to find patterns in the surviving sequence $c_1, \dots, c_n$ and take decisions how to instruct things further. As we 'll see immediately after, there is no reason to describe in detail such an algorithm though. We just adopt the mere assumption that the procedure generally works as follows:

Assuming that such a pattern can be intentionally constructed by applying successively the algorithm of an executable code $r$ to $c_k$ to obtain $c_{k+1},$ $k< n,$ we can search for such an $r,$ by defining an algorithm $\Delta$ that calculates in a row codes $r$ and checks if they {\em fit} the sequence, i.e. it checks if for every $k < n$ it is true that applying the algorithm of the code $r$ in question to $c_k$ has as a result the corresponding value in $c_{k+1}.$\footnote{Regarding the case of non-convergence of $\alg(r)(c_k),$ see below.} One should have in mind that this relation is not necessary an equation, since $r$ may output in only a specific part of the codes. We say then that the code $r$ {\em fits} the transition $c_k \mapsto c_{k+1}$ and denote it as $\alg(r)(c_k) \sqsubseteq c_{k+1}.$ For example, in the first case above, $r$ could be an executable program that adds 1 to the specific part of the input code, whereas in the second case, $r$ could be an executable program that copies the specific parts that remain intact. 

In the case that $r$ fits either all the sequence or a recent part of it, $\Delta$ takes an appropriate decision to apply it for all (or for all but few) descendants of $c_n.$ We are going to use a specific term for such a procedure, namely {\em diagonalization.}\footnote{The reasons for this name are historical and go back to the original self-referential proof of Cantor, mentioned earlier.}

Assume now that $\delta = \delta_n$ is a code for $\Delta.$ By the potential capability of a self-editing code to retain its self-memory, i.e. (3) above, $c_n$ can also have as an input the all sequence $c_1, \dots, c_n,$ so that by the basic self-editing principle, if $c_n[\overline{\delta}_n]$ is the state of $c_n,$ then it will produce the same result as $\Delta$ with the additional and important feature that now $\Delta$ can {\em find and perpetuate also patterns of evolution of itself.} Indeed, assuming that $\delta_k,$ $k<n,$ are versions of $\delta$ as subcodes of $c_k = c_k[\delta_k]$ respectively, any pattern recognizable by $\Delta$ and contained in the seqeunce
\[ \delta_1, \delta_1, \dots, \delta_n,\]
will be perpetuated as well. The benefit is of course that $\Delta$ in this case, not only evolves the rest of the code, but also evolves its own code as well.

Consequences of this simple approach, are somewhat surprising and we just name a few here:

\smallskip
--It is well known by Turing's Halting problem that the relation \[\alg(r)(c_k) \sqsubseteq c_{k+1}\] is {\em undecidable.}\footnote{i.e. there is no general way to decide if a given  executable program with a given input will stop eventually and give an answer: It may as well run for ever.} Based on its experience and using diagonalization, a self-editing code can fix an upper time limit about how much it should wait for an answer. In general (much like us) it can deduce conclusions about undecidable relations based on a (finite) experience. (Example 3 and the remark immediately after, in page \pageref{undecidability}).

\smallskip
--Assuming that a self-editing code has a feedback for the success of its interaction with the environment, it can learn using diagonalization to select its successful answers to the external world and diagonalize upon them also. (In addition to diagonalizing upon all the surviving sequence). (Section \ref{subsequenceDiagonalization}, page \pageref{subsequenceDiagonalization}).

\smallskip
--Using diagonalization, a self-editing code may learn by experience to recognize first the most common  patterns, even if it hasn't been programmed to do so (Example 2 in page \pageref{searcherTraining}). Moreover, it can do so relatively to the task at hand (Learning specialization in page \pageref{learningSpecialization}).

\smallskip
--It can regulate its internal parameters, including the parameters of the diagonalization procedure, for example its memory, i.e. the length of the sequence that uses to perform diagonalization, relatively to the decision that it should make (Mental experiments 1 and 2 and Remarks 2 and 3 immediately after, page \pageref{mentalexperiment1}). 

\smallskip
--Assuming an internal representation of the environment (for which we are going to talk in section \ref{InternalRepresentation}, page \pageref{InternalRepresentation}), a self-editing code may using diagonalization to understand and predict smooth environmental changes, for example {\em velocity} and {\em acceleration}. (Mental experiments 1 and 2 and Remark 4 immediately after, page \pageref{mentalexperiment1}).

\medskip
The paper is organized as follows: Section \ref{codes and addresses} is devoted to basic notations and definitions that are used.
Sections \ref{algorithmicComputations} up to \ref{decisions} serve as a detailed approach to the general study of self-editing computations and their relationship with usual computing. Sections \ref{diagonalization} up to \ref{learningSpecialization} are the essential part of the study and are devoted on diagonalization and its capabilities relatively to learning. 
Section \ref{hints} is devoted to outline possible further research on the subject and also give some hints about it. Finally, section \ref{evidence} refers to related work on the subject and the evidence that is provided by it.

\medskip
{\bf Aknowledgements.} The research in this paper has been done in the course of many years and many people (mostly mathematicians) have contributed to it, in various ways. (In chronological order) Maria Avouri, James Stein, Despoina Zisimopoulou, Dimitris Apatsidis, Fotis Mavridis, Antonis Charalambopoulos, Vanda Douka, Antonis Karamolegos, Helena Papanikolaou and Miltos Karamanlis are among them.

\section{Structured codes and addresses} 
\label{codes and addresses} 
We will assume that an algorithm in general has both as an input and as an output {\em codes.}  A {\em code} is thought to be a string (or sequence) on a given (finite) set (or {\em lexicon}) $L$. For example codes on  $L = \{0, 1\}$ are the strings $001,$ $11001$ and so on. The codes that we are going to use are supposed to have a {\em structure,} something that offers the facility of either referring to them as {\em activated,} or use them as inputs or outputs. On the other hand, codes can serve also as descriptions of algorithms, which will be proved very useful in the sequel. In case we want to highlight the difference, we refer to such codes as {\em executable codes} or {\em executable programs}, while non-executable codes are referred to as {\em data codes}. Finally a code may contain both executable parts and data parts.

For reasons that will become apparent in the sequel, it is convenient to be able to refer to specific parts of a code, using specialized for this purpose algorithms. This procedure can be supported by the structure itself of a given code. As an example, we may consider the code
\[ 01 \;\; 0010 \;\; 0011\]
This code has three parts, the first one being the sub-code $01,$ the second one being $0010$ and finally the third one being $0011.$ This way one can refer to these three sub-codes using the symbols $(1),$ $(2)$ and $(3).$ 

Leaving blank spaces is a very common way to indicate structure and it just requires to assume that our lexicon $L,$ contains a blank space. However it is not at all convenient for mathematical analysis where structure is usually indicated by the use of left and right parentheses, namely by the symbols $($ and $).$ The advantage of using parentheses is that  one may use them as an easy way to indicate a nested structure of a code. As an example, we may consider the code
\[ ((01, 001),(11, 00), (111, (00, 11, 0))).\]

One can use also blank spaces to describe the same code:
\[ 01 \;\;\; 001 \quad\quad 11 \;\;\; 00 \quad\quad 111 \;\;\; 00 \; 11 \; 0,\]
yet this second way although more intuitive, induces ambiguities.
Traditionally, in mathematics, we can refer to the various parts of such a code, by a sequence of natural numbers indicating the position of the part being considered. For example, the sequence (1) may indicate the first of the three parts of the above code, namely $(01, 001),$ whereas the sequence $(3, 2, 2),$ indicates the second part of the second part of the third part of the code, which is $11.$ Sequences as the above ones that determine a part of a code will be called {\em addresses}.

In our exposition, we are going to replace parentheses with brackets: $[$ and $].$ The reason is not to confuse the structure of a code with the application of a function (usually algorithmic) that is applied on a code.  Moreover, for technical reasons, we prefer to use brackets to indicate in an unambiguous way a code written using blank spaces. This way, for example, the code $x$ is identified with the code $[x],$ something that will save us from time, space and unnecessary complexity during the study. 

On the other hand, we are going to preserve the above mentioned traditional way of addressing sub-codes, yet somehow extend it. We are going to use the term {\em address} not only for strict addresses explained above, but also for simple (algorithmic) functions that compute them. For example by address $(\last)$ we mean the last component of a code, i.e. the $n$-th part of a code of length $n.$ That is, the $(\last)$ address of $[c_1, \dots, c_n]$ contains the code $c_n.$

As explained also in the introduction, in order to describe that $b$ is a subcode of $c,$ we write $c$ as $c[b].$ The fact that $b$ is contained specifically in the address $\theta$ of $c,$ is denoted as $c[(\theta) b],$ where parentheses around $\theta$ in this case demonstrate that $\theta$ is not a part of the code. Similarly, we use the notation $c[(\theta) \varnothing]$ to indicate that $\theta$ is empty in $c$ and the notation $c[\varnothing]$ to describe a code $c$ with an empty address.

\section{Algorithmic computations} \label{algorithmicComputations}
In order to be able to study self-editing computations, we will have to describe a general way to compose algorithms, out of simpler ones, i.e. instructions. Although we talk about algorithms, we will retain the term {\em program} for such a description.

 Such an analysis consists of a finite sequence $B_1, \dots, B_n$ of elements of a set $\mathcal{B},$ (thought of as the set of {\em instructions}), plus an algorithm $B$ that controls the flow of the execution of $B_1, \dots, B_n.$ We assume therefore that $B_1$ is the first instruction to be executed and subsequently $B$ controls the next instructions based on the index of the last instruction executed and the computation up to this step. (This last one, since the flow of the execution may be conditional). More formally:

 A {\em $\mathcal{B}$-algorithmic computation} with {\em program} $(B: B_1, \dots, B_n),$ where \[B_1, \dots, B_n \in \mathcal{B},\] and $B$ is an algorithm that controls the flow, is described as follows:

Given a code $x$ (which serves as an input), $(B: B_1, \dots,B_n)$ is thought to produce a sequence of codes
\[ x = x_1\mapsto x_2 \mapsto \cdots \]
where:
\begin{enumerate}	
	 \item For every $k = 1, 2, \dots,$ there is an index $j_k$ in the range ${1, \dots, n},$ such that $x_{k+1} = B_{j_k}(x_k).$ In this case, $B_{j_k}$ is called the {\em activated instruction} of the {\em computational step} $x_k \mapsto x_{k+1}.$
	 \item We assume that $j_1 = 1,$ i.e. $x_2 = B_1(x_1).$
	\item For every $k = 1, 2, \dots,$ we have that $j_{k+1} = B(j_k, x_{k+1})$ i.e. the instruction that is currently used is an (algorithmic) function of the previous one and the current computation. (Notice that $x_{k+1} = B_{j_k}(x_k),$ so that $x_{k+1}$ is in fact the current computation).
\end{enumerate}

It should be clear that any computer program on any language is of the form $(B: B_1, \dots, B_n),$ where $B_1, \dots, B_n$ are instructions of the language and $B$ is a suitable algorithm that controls the flow of the execution. Conditions regarding the flow are thought to be contained in the computation $x_1, \mapsto x_2, \mapsto \cdots,$ which is the reason for letting $x_k$ to be part of the input of $B.$ For example, a loop-free program that executes the instructions $B_1, B_2, \dots, B_n$ in the same order, on input $x$ produces the computation \[ x= x_1\mapsto x_2 \mapsto \cdots \mapsto x_n \mapsto x_{n+1},\] where for all $k,$ $x_{k+1} = B_k(x_k)$ and is formalized by the algorithm $(B: B_1, \dots, B_n),$ where for all $k,$ $B(k, x_{k+1}) = k+1.$ (In this case, $B$ just follows the next instruction).

\medskip
{\bf Remark} By allowing each instruction of $B_1, \dots, B_n$ to be a program itself, i.e. a similar structure of the form $(C: C_1, \dots, C_m)$ (instead of being just an element in $\mathcal{B}$), we naturally arrive to the notion of a {\em structured  program}, that is a program that has nested programs as instructions, instead of elements in $\mathcal{B}.$ Clearly this procedure can be repeated, yielding in turn programs for $C_1, \dots, C_m$ and so on. In this case, $C_i,$ will be called sub-instructions of the original program. We will use the same terminology for the instructions of $C_i$ and so on.

 Structured programs enable us to use each of $B_1, \dots, B_n$ as arbitrary algorithms and be therefore able to analyze a computation to the desired degree.

\subsection{Proliferating computations} In what follows, we are going to abuse the usual notation about sets, by allowing elements in sets to be repeated. For example $\{x, x\},$ $x$
being a code, denotes a set of two identical codes and is thought to be different than $\{x\}.$ We should notice that this convention could be avoided, yet doing so would add a significant amount of unnecessary complexity in the present study. 

The concept of a {\em proliferating computation} follows by letting instructions in a program having multiple outputs. As a simple example, the function
\[ R: x \mapsto \{x, x\},\]
($x$ being a code) that having $x$ as an input, produces an output of two identical copies of $x,$ is clearly algorithmic. More generally, if $R_1, \dots, R_k$ is a set of algorithms, not necassarily different between them, let us denote by $\{R_1, \dots, R_k\},$ the algorithm defined as
\[ \{R_1, \dots, R_k\}(x) = \{R_1(x), \dots, R_k(x)\}.\]
Such an algorithm that outputs more than one codes will be called {\em proliferating}. The codes $R_1(x), \dots, R_k(x)$ are called {\em children} or {\em immediate successors} of $x,$ and the set that contains them is denoted by $\imsuc(x).$ The instruction $\{R_1, \dots, R_k\}$ is called the {\em activated instruction} of the step and similarly, in the (partial) computational step $x \mapsto R_i(x),$ $i=1, \dots, k,$ $R_i$ is called (as before) the {\em activated instruction} corresponding to the (partial) step. 

 As mentioned above, {\em proliferating programs} are programs with one or more proliferating instructions. Proliferating programs induce a branching computation, where each code may have more than one immediate successors. Such a structure is called a {\em tree,} and we usually denote it as $(x_t)_{t \in T},$ where the set of indices $T$ has the same tree-structure. Any element in a tree may have from $0$ to any finite number of {\em immediate successors} or {\em children}. The input $x = x_\emptyset$ is called the {\em root} of the tree and the conventions we make about this branching computation are similar to the ones we made about the simple one. Namely, given a proliferating program $(B: B_1, \dots, B_n)$ and a code $x$ as an input, the computation is defined as the tree $(x_t)_{t \in T}$ such that 
\begin{enumerate}
	\item The root of the tree $x_\emptyset$ is the input $x.$	
	\item For every $t \in T,$ there is an index $j_t$ in the range ${1, \dots, n},$ such that the set of immediate successors of $x_t,$ $\imsuc(x_t)$ is equal to $B_{j_t}(x_t).$ \item We assume that $j_\emptyset = 1,$ i.e. $\imsuc(x_\emptyset) = B_1(x_\emptyset),$ 
	\item If $t'$ is a child of $t,$ then $j_{t'} = B(j_t, x_{t'}),$ i.e. the instruction that is currently used is an (algorithmic) function of the previous one and the current computation.
\end{enumerate}

A {\em branch} of a computational tree $(x_t)_{t\in T}$ is a sequence
\[ x_{t_1} \mapsto x_{t_2} \dots,\]
such that for any $k,$ $x_{t_{k + 1}}$ is a child of ${x_{t_k}}.$

\section{Self-editing computations} 

Let $c = c[b]$ be an executable code that contains $b$ as an instruction in some address. 
Assume that $x_1, x_2, \dots, x_n, \dots$ is a computation of the algorithm that $c$ codes for, with input $x_1.$  Assume also that during a computational step $x_k \mapsto x_{k+1}$ of $\alg(c),$ $alg(b)$ is used as the instruction of the step. We say then that $b$ is the {\em activated code} of the computational step and that the code $c[b]$ is in {\em state} $c[\overline{b}].$ Codes in some state will simply referred to as {\em state-codes.} The {\em step-algorithm} of a state-code $c[\overline{b}]$ denoted as $\salg(c[\overline{b}])$ is defined to be the algorithm that $b$ (i.e the activated code) codes for:
\[ \salg(a[\overline{b}]) = \alg(b). \]

Assuming now that the sequence $c_1, c_2, \dots$ is a sequence of state-codes, a computation
\begin{equation} \label{self-editing.computations.1} c_1 \mapsto c_2 \mapsto \cdots \mapsto c_k \mapsto \cdots \end{equation}
is called {\em self-editing} if for every $k,$ $c_k$ plays both the role of the data-code of the computation and at the same time the role of the state-code that performs the computation. More formally, the computation \eqref{self-editing.computations.1} is called  {\em self-editing} if: 
\[ c_{k+1} = \salg(c_k)(c_k) \quad \text{for every $k = 1, 2, \dots$}.\]
Every calculating step $c_k \mapsto c_{k+1}$ of a self-editing computation will be called a {\em self-editing step} and will be abbreviated as $\selfed(c_k).$ I.e.
\[ \selfed(c_k) = \salg(c_k)(c_k) = c_{k+1}.\]

If $c_k \mapsto c_{k+1}$ is a self-editing step, both the state-code $c_k$ and its step-algorithm $\salg(c_k)$ will be called {\em self-editing}.

We can now state the main ingredient of our study which is the following:

\medskip
{\bf Basic Self-Editing Principle.} {\em For every algorithm $B,$ there is a code $b,$ (which is in fact a code for $B$) such that for every code $c[\varnothing],$ 
\[ \selfed(c[\overline{b}]) = B(c[\overline{b}]) \]}

\begin{proof} As stated, let $b$ be a code for $B$ and apply the definitions:
	\[ \begin{split} \selfed(c[\overline{b}]) = & \salg(c[\overline{b}])(c[\overline{b}]) \\
	 = & \alg(b)(c[\overline{b}]) \\ 
	 = & B(c[\overline{b}]) \end{split} \]
	 as needed.
\end{proof}

\subsection{Self replication} \label{selfReplication}
A first simple application of the Basic self-editing principle results considering as $B,$ the proliferating algorithm that computes two exact copies of its input. I.e., let $B$ be defined by
\[ B(x) = \{x, x\}, \quad \text{$x$ being any code}.\]
The basic self-editing principle then asserts that if $b$ is a code for $B,$ and for any code $c[\varnothing],$ the self-editing computational step of $c[\overline{b}]$ is a duplication of itself, i.e. (using the above set-like notation)
\[ \selfed(c[\overline{b}]) = \{c[\overline{b}], c[\overline{b}]\}.\]

Clearly, this self-editing computational step is (since $b$ remains active) by definition  repeated for each of the two descendants, creating a tree-like structure $(x_t)_{t \in T},$ where $T$ is the {\em dyadic tree}, (i.e. every node has exactly two children) and for all $t \in T,$ $x_t = c[\overline{b}].$

As we will see, more complex self-editing computations, generate trees with varying finite number of immediate descendants.

\medskip
It is clear that a proliferating self-editing computational step need not be just self reproducing in the sense that instead of producing copies, it may produce variations of itself, something that is going to be proved much more interesting regarding evolution. The general case of such proliferating steps may be stated as follows:

\medskip
{\bf  Proliferating principle} {\em Let $B_1, \dots, B_n$ be algorithms. Then there is a code $b$ such that for any code $c[\varnothing],$
\[ \selfed(c[\overline{b}]) = \{ B_1(c[\overline{b}]), \dots, B_n(c[\overline{b}])\}.\]

If moreover $b_1, \dots, b_n$ are codes for $B_1, \dots, B_n$ respectively, then we may assume that they are also sub-codes of $b.$}

\begin{proof} Let $B$ be the algorithm that having a code $c$ as an input computes:
	\[ B(c) = \{B_1(c), \dots, B_n(c)\}.\]
Let also  $b$ be a code for $B.$ By the Basic self-editing principle, we get that 
for any code $c[\varnothing],$
\[ \begin{split} \selfed(c[\overline{b}]) = & B(c[\overline{b}]) \\
= & \{ B_1(c[\overline{b}]), \dots, B_n(c[\overline{b}]) \}. \end{split} \]

For the second part of the statement, if $b_1, \dots, b_n$ are codes for $B_1, \dots, B_n,$ define the code $b = s[b_1, \dots, b_n]$ as a code for the following algorithm:
\textsf{  
\begin{enumerate} \item Let $w$ be the input
	\item For $i = 1, \dots, n$:
	\begin{enumerate} \item Activate $b_i$ on input $w$ and output the result $\alg(b_i)(w).$
	\end{enumerate}
\end{enumerate} 
}

Then clearly, for any code $c[\varnothing],$
\[ \begin{split} \selfed(c[\overline{b}]) & = \alg(b)(c[\overline{b}])\\ & = \{\alg(b_1)(c[\overline{b}]), \dots, \alg(b_n)(c[\overline{b}])\} \\ &= \{B_1(c[\overline{b}]), \dots, B_n(c[\overline{b}]) \}, \end{split} \]
as needed.
\end{proof}

{\bf Remark} Using the same notation as above, $B_i,$ $i = 1, \dots, n$ is called the {\em active instruction} of the (partial) step $c \mapsto B_i(c).$ In case that $b_i$ is both a code for $B_i,$ and a sub-code of $b,$ it will be called as the {\em active code} of the (partial) step $c \mapsto B_i(c).$

\medskip
Let us explore some other examples of the application of the proliferating principle, which also demonstrate the versatility of its use:

\medskip
{\bf Examples.} 1) \label{recursion} We may assume that a self-editing code contains addresses for input and output regarding to the environment. Such addresses will be denoted as {\em environmental input} and {\em environmental output} respectively, in order to distinguish them from the entire code which is considered both the input and the output of a self-editing algorithm. So let $[\overline{b}, 0]$ be such a code, where environmental output occurs in address $(2)$ and thus the code outputs $0$ in this case. According to the proliferating principle, $b$ can be obtained by defining two algorithms $B_{1}$ and $B_{2}$ as:
\[ \begin{split} B_{1}([\overline{b}, n]) & = [\overline{b}, n+1] \quad \text{and} \\
 B_{2}([\overline{b}, n]) &= [\overline{b}, n+2]. \end{split}\]
 
 Thus the self-editing computational step of $[\overline{b}, 0]$ gives two descendants that output to the environment $1$ and $2$ respectively:
 
\[ \selfed([\overline{b}, 0]) = \{B_1([\overline{b}, 0]), B_2([\overline{b}, 0])\} = \{[\overline{b}, 1], [\overline{b}, 2]\}.\]

By definition, this is repeated for each of the two descendants, so $[\overline{b}, 1]$ gives again two descendants that output to the environment $2$ and $3$ respectively, whereas $[\overline{b}, 2]$ gives two descendants that output $3$ and $4$ and so on. The situation can become much more complicated by defining $B_i,$ $i = 1, 2$ to act also on $\overline{b}.$  This can change the way that the descendants are calculated. The following is an example of this (permitted by self-editing) situation.  We are not going to use it further for the moment, until  section \ref{diagonalization}.

\smallskip
2) \label{selfAction} Let $\theta_1$ and $\theta_2$ be two addresses. Their {\em concatenation} is denoted as $\theta_1^\frown \theta_2,$ and is defined as the address $\theta_2$ of the code in address $\theta_1.$ For example if $\theta_1 = (1, 2)$ and $\theta_2 = (1),$ then $\theta_1^\frown \theta_2 = (1, 2, 1).$

Let $B(n, \theta)$ be the algorithmic function:
\[ B(n, \theta): a[(\theta) k] \mapsto \{ \underbrace{a[(\theta) k+1], \dots, a[(\theta) k+1]}_{\text{$n$ times}}\},\]
that is, $B(n, \theta)$ outputs $n$ identical codes by adding $1$ in $\theta$ address of the input code. 

Let $b[(\theta_1) n][\theta] = b[n][\theta]$ be a code for $B(n, \theta)$ and $c[(\theta_2)\varnothing] = c[\varnothing]$ be any code with empty $\theta_2$ address. Define $\theta_0 = \theta_2^\frown \theta_1.$ Notice then, that in the code $c\big[b[n][\theta]\big],$ $b[n][\theta]$ lies in the $\theta_2$ address and in $b[n][\theta],$ $n$ lies in the $\theta_1$ address. Therefore in $c\big[b[n][\theta]\big],$ $n$ lies in the $\theta_0$ address. Thus, adopting the notation
\[ n \times c = \underbrace{c, \dots, c}_{\text{$n$ times}},\]
we get that
\[ \alg(b[n][\theta_0])\Big(c\big[b[n][\theta_0]\big]\Big) = \{ n \times c\big[b[n+1][\theta_0]\big]\}, \]
which shows that 
\[ \selfed\Big(c\big[\overline{b[n][\theta_0]}\big]\Big) = \{n \times c\big[\overline{b[n+1][\theta_0]}\big]\}.\]
Concluding, the self-editing computational step of $c[b[n][\theta_0]]$ gives $n$ children with varying genetic behavior so as to compute $n+1$ children with varied genetic behavior so as to compute $n+2$ children and so on.

It is easy to see that one can alter this example to describe self-editing codes that compute children with varying genetic behavior among them. Let us call for the moment $n$ as {\em genetic variable} and let $[n/3]$ be the closest from below natural number to $n/3.$ By defining 
\[ B(n, \theta): c[(\theta:)k] \mapsto \{ [n/3] \times c[k+1], [n/3] \times c[k], [n/3] \times c[k-1]\}\]
one arrives similarly as above to define self-editing codes that give children of increased, the same and decreased genetic variables. Since in a less theoretical environment available resources limit the production of children, selection would act in this case, so that the branches with the more appropriate number of children would survive. Yet selection could not alter the original behavior described by $B(n, \theta)$ above. Even in surviving branches this strategy would remain the same. In section \ref{diagonalization}, we are going to introduce a general method called {\em diagonalization} by which a self-editing algorithm may learn to choose the number of children according to the available resources and alter thus the genetic behavior described by $B(n, \theta)$ more appropriately.

\subsection{Self-editing trees} 

The proliferating principle stated above induces the notion of a {\em proliferating self-editing computation}:

Given a state-code $c_\emptyset,$ a {\em proliferating self-editing computation} is a tree $(c_t)_{t\in T}$ with root $c_\emptyset,$ such that the children of every $c_t$ are defined by the application of $\salg(c_t)$  to the code $c_t,$ i.e. such that
\[ \imsuc(c_t) = \selfed(c_t) = \salg(c_t)(c_t) \quad \text{for every $t \in T.$}\]
For simplicity reasons, in the case that the step-algorithm of $c_t$ produces a single output, we identify the singleton $\{\selfed(c_t)\}$ with the code $\selfed(c_t)$ and call the corresponding step {\em non-branching}. So, a self-editing computational sequence (where a code may change its value via the algorithm that describes) may be thought of as a special case of a self-editing tree.

\section{Programming a self-editing tree}

As we saw in the basic self-editing principle, any external algorithm $B$ may be executed by a self-editing code internally. Assuming that $B = (C: C_1, \dots, C_n)$ is a program, we would like to examine if it is possible for a self-editing code to use the same calculations as does the program $B.$ This is particularly useful in the cases that we want to analyze a particular computation performed by a self-editing code into simpler ones.  

As we 'll see in the following {\em Programming lemma} the answer to this can be roughly stated as follows:

\smallskip
{\em For every algorithm $P,$ there is a state-code $p$ (which is roughly the executable program of $P$), such that for every code $c[\varnothing],$ the computation of $P$ with input $c[p]$ is the same as the self-editing computation of $c[p],$ providing that $P$ does not alter its code i.e. $p.$}

\smallskip
Indeed, if $P$ changes $p$ during the computation, the above statement cannot be expected to hold, since in this case the corresponding self-editing computation will change its own intended program by mimicking $P.$ Although we are going to appeal to this kind of phenomena later on (and will be proved the main targets for studying), for the time being we rather wish to avoid them. Therefore one should consider programs that during their computation do not alter the address of $p$ above. In this direction, we need the following: 

We will assume that the result of a program $P$ that outputs the code $s$ in address $\theta$ of the code $c[(\theta) b],$ is $c[(\theta) s]$ (and not just $s$), unless other algorithms output to different addresses of the code. (For this reason, we will write $P(c[b]) \sqsubseteq c[s],$ instead of $P(c[b]) = c[s]$ which might be misleading). For simplicity reasons, this convention is made for all possible immediate descendants of the code in the computational tree. After this convention, a program $P$ will be called $\theta$-stable if for every computational step $c_1 \mapsto c_2$ instructed by $P,$ the $\theta$ address of $c_1$ coincides with that of $c_2.$ Clearly if a program $P$ is $\theta$-stable then every instruction of $P$ that may be used in a computational step ought to be $\theta$-stable too.

Secondly, $P$ is not expected to activate the correct codes during the computation, something that is expected to happen during the self-editing computation of $c[p],$ by the definition of the states of $p$ during the computation. Thus one needs to compare the deactivated versions of the two computations, defined below:

Two state-codes are called {\em code-equivalent} if they are state-codes of the same code. Similarly two computations $(x_t)_{t \in T}$ and $(x'_t)_{t \in T'}$ are called {\em code-equivalent} if $T = T'$ and for every $t \in T,$ $x_t$ is code-equivalent to $x'_t.$

Granting the above definitions and notations, we can prove the following (where the code in the stable address is exactly $p$):

\medskip
{\bf Programming Lemma} {\em Let $P$ be a $\theta$-stable (proliferating) program. Then there is a state-code $p,$ such that for any code $c[(\theta)\varnothing],$   the self-editing computational tree of $c[(\theta) p]$ is code-equivalent to  the computational tree of $P$ with input $c[(\theta) p].$ 
	
	Moreover, if $P = (B: B_1, \dots, B_n),$ and $b, b_1, \dots, b_n$ are codes for the algorithms $B, B_1, \dots, B_n$ respectively, then we may assume also that $b, b_1, \dots, b_n$ are sub-codes of $p.$}

\smallskip
Before passing to the proof, for $\theta$ an address and $i$ a natural number, let us denote by $\theta^\frown i$ the address that results by appending $i$ to $\theta.$ For example, if $\theta = (1, 3),$ $\theta^\frown 4 = (1, 3, 4).$

\begin{proof}
	For the proof we are going to code both the function of $B_i$ and of the flow-controlling function $B,$ in a code $s_i =s[b, b_i]$ described below. Notice, that although $P$ is supposed to be $\theta$ stable, the algorithms of $s_i$ shouldn't be, since they have to inactivate and activate the appropriate instructions.	Notice also that for any code that contains $[s_1, \dots, s_n]$ in its $\theta$ address, the addresses $\theta^\frown i,$ $i=1, \dots, n,$ correspond to the codes $s_i$ respectively. In the following algorithm we assume that the input is always a code that contains $[s_1, \dots, s_n]$ in its $\theta$ address and in a state that one of $s_1, \dots, s_n$ is activated, something that can be proved afterwards by induction.
	
	So, let $s_i =s[b, b_i]$ $i =1, \dots, n,$ be a code for the following algorithm:
	\textsf{
		\begin{enumerate} \item Let $w$ be the input. {\normalfont\textit{(We assume here that $w$ is of the form $x[(\theta^\frown i) [\overline{s[b, b_i]}]].$)
			}}
			\item Let $\{x_1, \dots, x_k\}$ be the output of activating $b_i$ with input $w.$ {\normalfont\textit{(Notice that by assumption $b_i$ is a code for $B_i$ which is $\theta$ stable)}}
			\item	For every $j=1, \dots, k:$ {\normalfont\textit{(i.e. for every code generated by $B_i$)}}
			\begin{enumerate} \item Let $i'$ be the output of activating $b$ with input $(i, x_j).$ {\normalfont\textit{(Since $b$ is a code for $B,$ this should give the number of the next instruction to be executed).}}
				\item Let $x'_j$ be the result of deactivating the code in address $(\theta, i)$ of $x_j.$ {\normalfont\textit{ (The particular code should be $\overline{s[b, b_i]}.$)}}
				\item Let $x''_j$ be the result of activating the address $\theta^\frown i'$ of $x'_j.$ {\normalfont\textit{(So that the correct instruction is activated).}}
				\item Output $x''_j.$
			\end{enumerate}
		\end{enumerate} 	}
	
	Set $p = [\overline{s}_1, s_2, \dots, s_n].$ Let $(x_t)_{t\in T}$ be the computation of $P$ and $(y_t)_{t\in T'}$ be the self-editing computation starting with $c[p].$ 
	
	Notice first that since $P$ is $\theta$-stable, for every $t,$ the code in address $\theta$ of $x_t$ must be $p.$
	
	Assume that for some $t \in T,$ it is true that $t \in T',$ $x_t$ and $y_t$ are code-equivalent and that if $B_i$ is the active instruction of $P$ then $s_i$ is activated in $y_t.$ (Notice that this assumption is true for the roots). So the immediate successors of $y_t$ are the result of $\alg(s_i)(y_t).$ 
	
	Since $B_i$ and $\alg(s_i)$ have the same number of output codes (which is $k$ in the algorithm described above), it follows that the immediate successors of $t$ in $T$ are the same as the immediate successors of $t$ in $T'.$ On the other hand, for every $x_j$ that $B_i$ outputs, $\alg(s_i)$ outputs a state-code $x''_j$ which results from deactivation and subsequently activation processing of $x_j.$ (See the instructions (b), (c) and (d) above). Therefore $x_j$ and $x''_j$ are state-codes of the same code. Furthermore $P$ activates the instruction with index $i' = B(i, x_j)$ for $x_j$ and in $x''_j$ is activated the code in address $\theta^\frown i'$ which should be the $i'$-component of $p,$ therefore $s_{i'}.$ This concludes the inductive step and therefore the proof.
\end{proof}

Though Programming lemma is necessary to state towards a more thorough understanding of self-editing, its most interesting applications as mentioned above, occur in the cases that $P$ is not $\theta$-stable, in which case $\alg(p)$ constructed in the proof, might change $p$ also. In this case there is no real ``program" behind the self-editing procedure that takes place, yet it gives the opportunity of evolving $P$ as we will see.

\section{Memory of self}
\label{selfMemory} Let $(c_t)_{t \in T}$ be a computation and fix a $t \in T.$ The {\em history} of $c_t$ is defined to be the sequence 
\[c_{t_1} \mapsto c_{t_2} \mapsto \cdots \mapsto c_{t_n} = c_t,\]
where $c_{t_1} = c_\emptyset$ is the root of the tree, and for every $i < n,$ $c_{t_{i+1}}$ is a child of $c_{t_i}.$ (Notice that $c_t$ itself is contained in this sequence). Below we denote by $\hist(c_t)$ the history of $c_t.$ 

Given a state-code $c = c_\varnothing,$ a {\em complete memory} self-editing computation, is a tree $(c_t)_{t \in T}$ such that for every $t \in T,$ 
\[ \imsuc(c_t) = \salg(c_t)(\hist(c_t)), \]
i.e. every state-code $c_t$ computes its children by applying its algorithm, not only on its code, but on the entire sequence of its history.

On the other hand, a {\em complete memory instruction}, is thought to be an algorithm $B,$ that computes the children of a code $c_t$ having as an input the history of $c_t,$ and a {\em complete memory} program is a program (as we have defined it) that uses complete memory instructions.

It is easy to see that both the basic self-editing principle and the programming lemma can be stated and proved in the case of complete memory algorithms. Below,we state the complete memory version of the basic self-editing principle for future reference: 

\smallskip
{\bf Basic self-editing principle.} {\em (Complete memory version) Let $B$ be an arbitrary algorithm with code $b.$ Let also $c_1, \dots, c_n$ be the history of the code $c_n$ in a computational tree. If the code $b$ is activated in $c_n,$ then the self-editing computation of complete memory performed by $c_n,$ is equal to $B(c_1, \dots, c_n).$
}

\smallskip
What is interesting here, is that although at a first glance, a self-editing computation of complete memory seems stronger than a simple one, this is not true. The idea behind this, is that a complete memory self-editing computational step can be simulated by a simple one that stores memory.

Fix $\phi$ to be an algorithmic correspondence with an algorithmic inverse between sequences of codes and codes. A simple such function is for example
\[ \phi : (c_1, \dots, c_n) \mapsto [c_1, \dots, c_n]. \]

Let us notice that, although for practical applications, this particular correspondence is not at all space-saving efficient, it fits very well a theoretical approach, so we are going to use it in what follows.

Hereafter, we are going to use $H(c_n) = \phi(\hist(c_n))$ to denote briefly $[c_1, \dots, c_n],$ where $c_1, \dots, c_n$ is the history of $c_n.$
Also, the address $(\last)$ of $\phi(\hist(c_n))$ is defined to be where $c_n$ (the currently active code) is lying.

Notice that by replacing a code $x$ with $\phi(\hist(x)),$ we assume that all addresses except the address $(\last)$ are memory addresses and thus inactive, even if they do have active components. The reason behind this approach is the obvious one: We would like to keep the remembrance of an active component, and on the other hand, introducing a new symbol for past activity would result in unnecessary complexity. Having this convention in mind, we arrive at the following definition:

A self-editing computation is called {\em memory storing} if it is of the form:
\[ [c_1], [c_1, c_2], \dots, [c_1, c_2, \dots, c_n].\] 
It is useful to notice, that as we have mentioned in section \ref{codes and addresses}, the initial code $[c_1]$ may be thought to be the same as $c_1.$ Before passing to state and prove that a self-editing computation of complete memory can be simulated by a simple one that instead stores memory, let us discuss the idea behind this, which is quite simple. Indeed, what we have to do is to replace a self-editing computational sequence
$c_1, \dots, c_n$ of complete memory with the corresponding one that stores memory, namely
\[ c_1, [c_1, c_2], \dots, [c_1, c_2, \dots, c_n]\]
and make sure that the algorithm that is described by the state code $c_n$ should be executed upon the sequence that is described by its code, i.e. $c_1, \dots, c_n.$ This of course would yield easily the result and it could be accomplished by the simplified instruction (where {\em appending} $x$ to $[c_1, \dots, c_n]$ is thought to result in the code $[c_1, \dots, c_n, x]$):

\smallskip
\begin{center} \textsf{\begin{tabular}{p{11cm}} If $[c_1, \dots, c_n]$ is the input, execute $c_n$ with input the sequence $c_1, \dots, c_n,$ append the result (or the results) to $[c_1, \dots, c_n]$, and output the computed code (or codes).\end{tabular}} \end{center}

\smallskip
If $m$ is a code for such an instruction and we replace each $c_k$ in the sequence by $c_k[\overline{m}],$ then the result follows easily and immediately. However, in practice this procedure meets certain technical difficulties. On the other hand, analyzing and solving them, has the advantage that one discovers a great deal of the surprising versatility of using self-editing. So let us first indicate and discuss them:

\smallskip
1) First of all, the code $m$ in $c_n[\overline{m}]$ should be of a greater priority than any activated code in $c_n$ alone. This has to be so, since the activated codes in $c_n$ should be stopped from being executed at first place, until the sequence $c_1, \dots, c_n$ which serves as an input to $\salg(c_n)$ is computed from $[c_1, \dots, c_n].$ So we will have to use some notion of priority to the codes that are being executed, something that we discuss immediately after. Yet, here comes another problem. To assign a priority to $m$ that is greater of all priorities used in the (infinite) sequence $c_1, c_2, \dots.$ This of course is impossible, yet self-editing as we will see, permits us to define $m$ so that $\alg(m)$ fixes its own priority in every step $c_k[\overline{m}] \mapsto c_{k+1}[\overline{m}]$ so that it exceeds that of every activated code in $c_{k+1}$ alone.
	
\smallskip
2) The second technical difficulty comes from the fact that we should be able to find an address to accommodate $m$ which is not an address that is used or being created during the  computational sequence. The easy (yet a bit technical) solution that we use here, is to ``add a dimension" in the structure of the code by letting $c_n[\overline{m}]$ to stand for $[[c_n], \overline{m}].$ 

\smallskip
Let us mention here that the notion of {\em priority} is met both in biology, where a gene can stop another one from being expressed, and equally well in neuroscience, where a neuron can stop another one from firing. We will model it here, by assuming that in some relative address of an instruction, say $\theta_p$ lies a natural number that determines the ``strength" of the instruction relative to other ones.

\smallskip
Since the notation $c[\overline{m}]$ has been reserved to symbolize that the (active) code $m$ is a subcode of $c,$ we use $c[(+) \overline{m}]$ to denote that a new address has been appended to $c$ that contains $\overline{m}.$

We state and prove at this point the relevant result:

\medskip
{\bf Memory lemma.} {\em Let $(c_t)_{t \in T}$ be a self-editing computation of complete memory. Then there is a code $m$ such that $c_\varnothing[(+)\overline{m}]$ results in the simple (without memory) computation $(x_t)_{t\in T},$ where for every $t \in T,$ if $c_1, \dots, c_n = c_t$ is the history of $c_t$ in $(c_t)_{t \in T},$ then 
	\[x_t = [c_1[(+)\overline{m}], \dots, c_n[(+)\overline{m}]].\]
}

\begin{proof}
Let $m$ be a code for the following algorithm:

\begin{enumerate}
	\item \textsf{Let $x$ be the input.} {\em (We assume here that $x$ should be of the form $[[[c_1], \overline{m}], \dots, [[c_n], \overline{m}]]$ and moreover the priority of $m$ in each of the codes exceeds that of the corresponding $c_i.$).}
	\item \textsf{Stop any instruction from being executed in address $(\last, 1, 1).$} {\em (Notice that this address should contain $c_n$).}
	\item \textsf{Let $n$ be the length of $x.$} 
	\item \textsf{For $i = 1, \dots, n,$ let $z_i$ be the code in address $(i, 1, 1)$ of $x.$} {\em (So that in fact $z_i = c_i$).}
	\item \textsf{Execute the state-code $z_n$ with input $z_1, \dots, z_n$ and let $\{w_1, \dots, w_s\}$ be the result.} {\em (So that $w_1, \dots, w_s$ are in fact the children of $c_n$).}
	\item \textsf{Let $y$ be the code in address $(\last, 2)$ of the input.} {\em (So that $y$ should be equal to $m$ itself).}
	\item \textsf{For $j = 1, \dots, s$:}
	\begin{enumerate} \item \textsf{Fix the priority of $y$ to be strictly greater than that of any activated code in $w_j.$}
		\item \textsf{Append $[[w_j], \overline{y}]$ to the input $x$ and output the result.} 
	\end{enumerate}	
		{\em (Thus according to our initial assumption, the output should be all the codes of the form
		\[ [[[c_1], \overline{m}], \dots, [[c_n], \overline{m}], [[w_j], \overline{m}] ],\]
		where $w_j$ ranges over the children of $c_n$ and moreover the priority of $m$ in the above code exceeds the one of any activated code in $w_j.$}
		\end{enumerate}
Fix now the priority of $m$ to be strictly greater than that of any activated code in $c_\varnothing$ and let $x_\varnothing = [[c_\varnothing], \overline{m}].$ Thus the assumption we make in the instruction (1) of the definition of $m$ holds true for $x_\varnothing$ and the rest comments next to the instructions provide an inductive proof of the result.
\end{proof}

Memory principle allows us to consider a simple self-editing computation as one of complete memory, assuming that the memory is stored. Nevertheless, in a self-editing computation, memory storing is sometimes not needed. This is made obvious in the following examples:

\medskip
{\bf Examples.} Let $R$ and $S$ be algorithms. and $x_1, \dots, x_n =x$ the history of a code $x.$ Assume that for every $k=1, \dots, n,$ the computation up to $x_k$ is memory storing, i.e. there are codes $c_1, \dots, c_n,$ such that $x_k = [c_1, \dots, c_k].$ Given an algorithmic function $R,$ notice that the computations
\[ [c_1, \dots, c_n] \mapsto R(c_1) \quad \text{and} \quad [c_1, \dots, c_n] \mapsto [c_1, \dots, c_n, R(c_1)],\]
are both algorithmic, so that by the basic self-editing principle, if their codes are existent and activated in $c_n,$ will produce them. The main difference is that while the second one continues to store memory, the first one deletes it. We are going to refer to both of them as {\em cycles}, since they are thought to produce a variation (or even a copy for $R$ being the identity), of the initial code.

Another interesting example of this kind, is the proliferating self-editing computation
\[ [c_1, \dots, c_n] \mapsto \{R(c_1), [c_1, \dots, c_n]\},\]
which may be thought of as modeling (for $R$ being the identity) the self-replication of multi-cellular organisms (with one parent).

\section{Possible decisions of a self-editing code} \label{decisions}
\subsection{Storing decisions.} \label{storingDecisions} Let $r$ and $c$ be codes and let us consider the following algorithmic function $B(r)$ that acts having as input the code $c$:
\begin{equation} \label{avAd} B(r): c \mapsto \begin{cases} [c_1, \dots, c_n, r] & \text{ if } c = [c_1, \dots, c_n],\\
[c, r] & \text{ if not } \end{cases}
\end{equation}
which simply appends $r$ on $c.$ The {\em basic self-editing principle} ensures us that if $c$ is self-editing, $b = b[r]$ is a code of $B(r),$ and $c = c[\overline{b}]$ is the corresponding state of $c$ in which $b$ is activated, then the resulting self-editing step of $c$ is the one described by \eqref{avAd}. Notice that the address that is used by $B(r)$ to store the code $r,$ is a {\em new} one, i.e. one that was not existed previously in $c.$ Such an address will be called an {\em available address}, so that we can abbreviate the corresponding self-editing step of $c[\overline{b}],$ as resulting from the instruction:

\smallskip
\textsf{\begin{tabular}{p{10cm}} Store $r$ in an available address. \end{tabular}}

\smallskip
To denote that $\theta$ is an available address in a code $c,$ we will write that $c = c[(+\theta) \varnothing]$ or in a more simple form $c = c[(+)\varnothing].$ (Notice that in this case $\theta$ is not existent in $c,$ yet it may be created).

\subsection{Temporary decisions.} Let $r$ and $c$ be codes and $\theta$ an available address in $c.$ Let also $s=s[r]$ be a code for the algorithm that is described by (For the reader's convenience the input is thought to be $c[ (+\theta) \overline{s}]$ ):

\smallskip
\textsf{\begin{tabular}{p{10cm}}
		Let $x_1$ be the input.\\
		Let $x_2$ be the result of deleting the address $\theta$ from $x_1.$\\
		Output $\alg(r)(x_2).$	
\end{tabular}}

\smallskip
It can be easily checked that the self-editing step of $c[(+\theta) \overline{s}]$ results in $\alg(r)(c).$ The code $r$ in this case will be called a {\em temporary differentiating code}.

It is clear that a temporary differentiating code may be invoked by a self-editing algorithm:

If $b = b[r]$ is a code for the calculation:
\[ c[(+\theta) \emptyset] \mapsto c[(+\theta) s[r]] \mapsto \alg(r)(c),\]
then by the basic self-editing principle the self-editing step of $c[\overline{b}]$ would result in the same calculation. We may abbreviate this step by saying that $c$ follows the instruction:

\smallskip
\textsf{Use $r$ as a temporary differentiating code.}

\subsection{Permanent decisions.} \label{permanentDecisions} Let us, as before, consider $r$ and $c$ to be codes and $\theta$ an available address in $c.$  We define $s= s[r]$ to be a code for the following algorithm (Again for the reader's convenience the input is thought to be $c[(+\theta) \overline{s}]$):

\smallskip
\textsf{\begin{tabular}{p{10cm}}
		Let $x_1$ be the input.\\
		Let $x_2$ be the result of activating $r$ with input $x_1.$\\
		Activate the code in address $\theta$ of $x_2$ and output the result.	
\end{tabular}}

\smallskip
One can easily check as before, that the self-editing step of $c[(+\theta) \overline{s}]$ results in a code $c_1$ that is computed by activating the $\theta$ address of $\alg(r)(c[(+\theta) \overline{s}]).$ Thus, assuming that $\alg(r)$ is $\theta$-stable, $c_1$ should be of the form $c_1[(\theta)\overline{s}].$ So by induction, the $n$-th step should result in $(\alg(r))^n(c[(+\theta) \overline{s}])$ by repeating $n$ times the application of $\alg(r)$ to $c[(+\theta) \overline{s}].$ The code $r$ in this case, will be called a {\em permanent differentiating code}.

Again, it is easy to see, that exactly as in the case of temporary differentiating codes, the application of a permanent one, may be invoked by a self-editing algorithm itself. We are going to abbreviate such an instruction by:

\smallskip
\textsf{Use $r$ as a permanent differentiating code.}

\smallskip
{\bf Remark.} Theoretically there is not  much difference between a temporary decision and a permanent one: The first can be repeated continuously, while the second may be interrupted by a code of greater priority. There is much difference in practice though, since a continuously used temporary differentiating code, costs the repeated complexity of the decision to use it.

\subsection{$\phi$-differentiating decisions} Given a self-editing code $c,$ an address $\phi$ in $c$ and a differentiating code $r,$ a temporary (or permanent) decision to differentiate $\phi$ by $r,$ is the decision whose temporary (respectively permanent) differentiating code is the code $s = s[\phi, r]$ of the following algorithmic function:
\[ c = c[(\phi) y] \mapsto c[(\phi) \alg(r)(y)],\]
which replaces the content $y$ of $\phi,$ by $\alg(r)(y).$
 The address of $r$ in $c$ will be called the {\em address of the differentiating code of $\phi.$}

\section{Diagonalization} \label{diagonalization}

The combination of the complete memory version of the Basic self-editing principle and the Memory lemma, allow us to consider a simple self-editing code that retains its memory as a self-editing code of complete memory. Henceforth, to avoid unnecessary confusion, we are going to study self-editing computations of complete memory, having in mind that our models are ones that instead store it, so that we may freely apply the complete memory version of the Basic self-editing principle. As in the simple case, it would be helpful for the reader, to freely translate this principle here, as follows:

{\em Assume that by looking at the history $c_1, c_2, \dots, c_n$ of a self-editing code, we decide to change the programming of $c_n$ in a particular way that can be expressed as an algorithm. Then the same decision can be potentially made by $c_n$ itself, assuming that its computation is self-editing and the memory of its history has been stored.}

The theory of diagonalization applies both to {\em surviving} and {\em successful} sequences.

While a surviving sequence can result directly from natural (or artificial) selection applied upon a tree of self-editing codes, a successful sequence results from an internal selection of successful computations. (In section \ref{subsequenceDiagonalization}, page \pageref{subsequenceDiagonalization} we examine how such an internal selection may result as a consequence of diagonalizing over the surviving history of a self-editing code). 

Assume that we have the information of an entire successful or surviving sequence and we ask ourselves what we would do if we were to help the evolution of such a code and yet we knew nothing either about the selection that is performed upon it, or about the true functionality of its code or of parts of it. (Notice that this is about the same knowledge that we assume that our code really has at the beginning). This is certainly not an easy task, yet the question has an easy, simple and natural answer: Since the successful or surviving sequence manifests by itself what is acceptable by the environment, either as a reward or as a selection, we should look for patterns in the code, aiming to perpetuate them. Let us formalize this a bit more: Assuming that the sequence \[c_1 \mapsto c_2 \mapsto \cdots \mapsto c_n\]  is either a successful one, or a surviving branch of the computational tree of a self-editing code, a {\em pattern} could be thought of as an algorithm $R,$ such that for every $k<n,$ $R(c_k) \sqsubseteq c_{k+1}.$ Several examples should clarify this concept. Suppose that we notice that a particular part of the code remains the same. Assume moreover that the position of this part is described by an address $\theta.$ This observation is equivalent for us to note that the algorithm $R$ defined by

\begin{center}
	\textsf{Copy the $\theta$ address of the input to the $\theta$ address of the output,}	
\end{center}

{\em fits} the sequence, i.e. for every $k< n,$ $R(c_k) \sqsubseteq c_{k+1}.$ This way we could advice or program the code to start applying it on purpose. (Notice that the validness of this rule up to now  upon the sequence, could be either a product of selecting a successful sub-sequence, or a product of selection). Therefore from now on, $R$ will be applied on purpose. It is easy to see that one could find a lot of examples of this kind. Here is one more: 

Assume we notice that a part of the code contains a parameter in the form of a natural number, that is continually augmented by, say $1.$ If $\theta$ is the address of this part, this is equivalent for us to observe that the algorithm $R$ defined by:

\begin{center} \textsf{Add $1$ to the $\theta$ address of the input and \\write the result in the $\theta$ address of the output,} \end{center}

fits the sequence.

Something useful to note here in order to grasp better the generality of this recognition of  patterns, is that our observation may as well be conditional, for example

\begin{center} \textsf{If it is true that $R(c_k)$ then compute $c_{k+1}$ by $S(c_k),$} \end{center}

where $R$ and $S$ are given algorithms. 

This procedure, of our advice which is based on our observation, can be roughly automated, using a pattern-recognition algorithm:

So, suppose that 
\[c_1[\delta_1], c_2[\delta_2], \dots, c_n[\overline{\delta_n}],\]
is an either surviving or successful sequence and let $\Delta$ be a pattern-recognition algorithm. Thus with the above sequence as an input, $\Delta$ should be able to recognize, (and taking the appropriate decision, to perpetuate), patterns of evolution of the sequence. At the same time, the same procedure can fulfill as well, the requirements of a smoothly changing environment, as they are probably imprinted in the structure of the codes of the sequence. (Mental experiments 1 and 2, in page \pageref{mentalexperiment1}, are examples of this). The main problem here, is the complexity of such an algorithm in order for it to function well. Our proposition, by means of self-editing, is essentially a simple algorithm that can evolve also itself: Indeed, assume that $\delta_n$ is a code for $\Delta,$ and $\delta_1, \dots, \delta_n$ the history of its evolution. Assuming Darwinism, this sequence should be of increasing effectiveness. Using then the complete memory version of the Basic self-editing principle and the Memory lemma, the next self-editing computational step of $c_n[\overline{\delta_n}]$ should be the same as the application of $\Delta$ to the sequence. This means that $\Delta=\alg(\delta_n)$ is able to recognize not only evolutionary patterns in $c_1, \dots, c_n,$ but also in $\delta_1, \dots, \delta_n,$ i.e. to itself. As mentioned in the Introduction, this relatively simple approach has surprising consequences, which we are about to study. 

Notice that our main target at this point, (yet not the only one), is to show that such an intelligent evolution could start almost without any knowledge.

\subsection{Diagonalizing algorithms} {\em Sequential diagonalization} is in fact a very general definition of an algorithm that finds patterns in a sequence. We do not aim here to give a detailed description of such a process. Instead, the description concerns any such algorithm, even not successful ones. The reason is that as it has been mentioned and will be made clearer by the examples, such an algorithm combined with the self-editing property, has the potential to evolve itself.

By the term {\em decision system}, we mean any algorithm $\Delta$ with the following general description:

$\Delta$ is used to make a decision $D$ and it consists of two parts, the {\em searcher} $\Delta_s,$ and the {\em tester} $\Delta_t.$

The searcher $\Delta_s,$ is thought to propose codes $r_1, r_2, \dots$ in a row. We will refer to these codes as the {\em proposed codes.} The set of proposed codes may be either finite, or infinite. For the infinite case, we assume that the searcher is designed so as to construct $r_1, r_2, \dots,$ by means of an initial finite set of instructions and an also finite set of ways of composing new codes out of old ones. There are general methods of such a construction, the more simple and less effective and sophisticated being to output arbitrary strings on a given alphabet. 

The {\em priority} of a proposed code is thought to determine its appearance in the sequence $r_1, r_2, \dots$ and we say that a code is {\em simple} if its priority is high, i.e. if it appears early in the sequence.

\smallskip
The tester $\Delta_t,$ is thought to use an algorithmic test $T$ (i.e. an algorithm that outputs \textsf{true} or \textsf{false}) in order to choose the simpler one of the proposed by the searcher code $r$ such that $T(r) = \textsf{true}.$ We will call $T$ the {\em testing algorithm} of $\Delta_t.$

\smallskip
Clearly, there can be a lot of variations of the above definition. For example, assuming that natural selection functions using a given algorithmic test $T,$ the system of a tree produced by a self-editing code and the environment, demonstrates a very general such variation. Notice that in this particular example, the proposed by the searcher sequence has been replaced by a self-editing tree. Nonetheless, for reasons of simplicity and for the moment, we will restrict ourselves to the case of proposed sequences rather than trees.

Another variation of this definition that should be kept in mind, is the case of a testing algorithm that outputs an evaluation instead of a (simplified for our purposes) true-false answer.

\smallskip
Given now an algorithm $S$ that inputs the history of a self-editing code $c$ and outputs a selected sub-sequence $c_1, c_2, \dots, c_n = c$ of it, we may consider the following testing algorithm $T:$

\smallskip
\textsf{\begin{tabular}{p{10cm}}
		Let $c_1, \dots, c_n$ be the output of $S.$
		Choose the simpler code $r$ in the proposed by $\Delta_s$ sequence $r_1, r_2, \dots$ of codes, such that $r$ fits the sequence $c_1, \dots, c_n,$ i.e.
		\[ \textsf{For all $i < n,$} \quad \alg(r)(c_i) \sqsubseteq c_{i+1}\]
		and use it as a temporary or permanent differentiating code.
\end{tabular}},

\smallskip
(For the convergence problem of such a test, we refer to Example 3 of page \pageref{undecidability}). The term {\em sequential diagonalization} is used to describe the functioning of a decision system $\Delta = (\Delta_s, \Delta_t),$ such that $\Delta_t$ uses the above test $T.$ The sequence $c_1, \dots, c_n$ will be called the {\em testing sequence} of the diagonalization.

The following principle is essential for our purposes:

\smallskip
{\bf Sequential diagonalization principle (Simple form)} {\em Let $\Delta$ be a sequential diagonalizing algorithm with code $\delta = \delta_n$ and $c = c[\overline{\delta}]$ a self-editing code. Assume that  \[c_1[\delta_1], c_2[\delta_2], \dots, c_n[\overline{\delta}_n] = c[\overline{\delta}] \] is the history of $c.$ Then
	\[\selfed(c[\overline{\delta}]) = \Delta(c_1[\delta_1], c_2[\delta_2], \dots, c_n[\overline{\delta}_n]). \]}

\begin{proof} It is an immediate consequence of the complete memory version of the Basic self-editing principle and the Memory lemma.
	\end{proof}

Notice that as it has been mentioned before, a diagonalization procedure in the frame of self-editing, may detect evolutionary patterns in its own program and perpetuate them. This could be a pattern in the sequence $\delta_1,\delta_2, \dots, \delta_n$ or in sub-codes of it. 

 It should be also noticed here, that ignoring the terms {\em surviving} and {\em successful}, diagonalization is in agreement with {\em Hebbian theory} \cite{hebb}. (For more details on this subject, see section \ref{hebbianPlasticity}, page \pageref{hebbianPlasticity}).

Before proceeding to study it better, let us see it in action, by means of some {\em mental experiments}. In these experiments, we are going to ask by a self-editing code to fill the dots in given sequences. Due to the lack of communication at this stage, instead of providing the known parts of the sequences, we are going to assume that during the self editing procedure they are somehow guessed no matter by which method.  Studying the mental experiments that follow, we 'll refer to this convention as {\em the basic guessing convention} and to such digits that we ought to provide as {\em not intended to be guessed digits}. Another one convention that we are going to make, is to assume that the set of candidate differentiating codes is well formed, that is, simple candidate differentiating codes for us, are hypothetically also simple for the self-editing code. (In Example 2 of page \pageref{searcherTraining}, we will see that due to its particular nature, diagonalization in the frame of self-editing, combined with adequate experience, may serve to the purpose of learning to detect useful codes and render them as simple).

\smallskip
{\bf Mental experiment 1.} \label{mentalexperiment1} In this initial mental experiment we ask by a self-editing code to fill the dots in $0, 1, 2, ....$ Of course we are going to assume as we have already done, an address to output to the environment, called {\em environmental output} and denoted by $\theta_e.$ 

Notice that it is necessary to provide the digits $0, 1, 2$ in order to guess the rest, therefore we may assume, due to the basic convention done above, that in a sequence $c_0, c_1, c_2$ the codes somehow guess the corresponding digit, i.e. for $i=0,1,2,$ $c_i = c_i[(\theta_e) i].$ Thus $c_0$ should write $0$ in its environmental output address, and similarly $c_1$ should write $1$ and $c_2$ should write $2.$

It is clear now, that the code $\textsf{add}(1)$ that adds one on a given integer is a $\theta_e$-differentiating decision that fits the sequence $c_0, c_1, c_2.$ Thus assuming that it is simple enough, it should be used as a temporary or permanent $\theta_e$-differentiating code. Therefore, if such a procedure as diagonalization is activated in $c_2$ the self-editing step $c_2 \mapsto c_3$ should yield $c_3$ so that $c_3 = c_3[(\theta_e) 3],$ that is $c_3$ should output $3$ to the environment. 

Clearly, the procedure can be carried on by induction, either by repeating diagonalization in the case of a temporary decision to use $\textsf{add}(1),$ as a $\theta_e$-differentiating code, or by the decision to use the same code permanently. Thus in either case, $c_3$ should yield $c_4 = c_4[(\theta_e) 4]$ and so on.

\smallskip
{\bf Mental experiment 2.} \label{mentalexperiment2} In this second experiment we aim to study the behavior of a self-editing code using diagonalization, on the subject of guessing the dots in a somewhat more complicated structured sequence, which is as follows:
\[ (0, 1, 2, \dots), (0, 2, 4, \dots,), (0, 3, 6, \dots), \dots.\]
That is, we ask not only to fill the dots in the first three sub-experiments, but also to guess entirely the fourth sub-experiment (as we certainly can do).

Guessing of the first three sub-experiments can be clearly accomplished the same way as mental experiment 1. An exception to this is the case of the closing right parentheses $),$ yet this is not a digit expected to be guessed. (Notice that we couldn't guess this digit also, since there is no indication where we should stop the successful guessing of the sequences of the sub-experiments). Thus right parentheses should follow the rule of the basic convention.

Notice now that right parentheses of the sub-experiments mark their ending,
Thus for example,

\smallskip
\textsf{\begin{tabular}{p{10cm}} Start every new sub-experiment with a left parenthesis followed by a $0,$ \end{tabular}}

\smallskip
can be regarded of as an alternative way of saying:

\smallskip
\textsf{\begin{tabular}{p{10cm}} If a left parenthesis is written in the address $\theta_e,$ then replace it in the next step with a right parenthesis and if a right parenthesis is written in $\theta_e$ then replace it in the next step with $0.$ \end{tabular}}

\smallskip
Clearly if $r$ is a code for such a statement, then $r$ should fit the memory sequence, since it has been followed up to now. 

Now notice that for $i= 0, 1, 2,$ during the $i$-th sub-experiment the address (let us call it simply $\theta$) of the $\theta_e$-differentiating code contains the code $\textsf{add}(i)$ respectively. Thus if $\theta_0$ is the (relative) address of $i$ in $\textsf{add}(i),$ then 

\smallskip
\textsf{\begin{tabular}{p{10cm}} In every new sub-experiment add $1$ in the address $\theta^\frown \theta_0$, \end{tabular}}

\smallskip
should fit the memory sequence as well, so that if $r$ is a simple enough code for it, by diagonalization it should be decided to be used as a differentiating code. Therefore a self-editing code should be able to guess that in the $4$-th sub-experiment the number-digits start with $0$ and are augmented by $4,$ thus being able to guess the entire sequence $(0, 4, 8, \dots).$

\smallskip
{\bf Remarks:} {\bf 1.} It is clear that we may continue this line of experiments by asking a self-editing code to guess a differentiating code for the differentiating code for environmental output, and so on. It is interesting to  mention here that the mere decision of creating a differentiating address for an existing one can be also a subject to be established by diagonalization. 

\smallskip
{\bf 2.} The environmental address $\theta_e$ is not an essential choice for the flow of the experiments and could be replaced by any address of the code. In fact both experiments, may be regarded as a convenient way to establish that a self-editing code should be able to improve its parameters based on an initial judgment about success. For a both simplified and interesting example of this, assume that $\theta$ is the address of the priority weight $w$ of a possible candidate differentiating code $r.$ A prejudged as successful sequence 
\[c[(\theta) w], c[(\theta) w+1], c[(\theta) w +2]\]
indicates that the weight should probably be augmented further and this is indeed the action that would be taken by diagonalization. It naturally remains open the problem of {\em homeostasis}, i.e. the problem of at which value an optimal (or locally optimal) is attained. This is going to be addressed later on, in Example 4, page \pageref{homeostasis}. 

Notice that the interest of the above remark, lies exactly onto that, assuming self-editing computations, diagonalization can practically be used to improve its own behavior (by regulating its own parameters),  as suggested somewhat theoretically earlier. For example it can be used to regulate the length of the testing sequence that is appropriate for a specific kind of decision. 

\smallskip
{\bf 3.} The potential ability of a self-editing code to decide the action to be taken in relevance to the length of memory that a given proposed code fits, can be seen also in a different manner, as suggested below:

Let $D(r)$ be any decision that concerns the proposed by the searcher code $r$ and $d$ be a code for $D.$ Then a self-editing code may randomly choose to apply $D(r)$ by activating $d[r],$ according to the basic self-editing principle. Assume also that for such a decision to be successful, it is required that $r$ fits at least $k_d$ of recent memory length. Thus in most cases of a surviving sequence (or successful subsequence, as we 'll see later) $c_1, \dots, c_n,$ it should be true that the decision $D(r)$ (which was made randomly) has been taken under the condition that $r$ should fit in at least $k_d$ length of recent memory. (We assume here that the probability of applying a correct decision $D(r)$ without the condition of $r$ fitting at least $k_d$ steps of recent memory is negligible). Therefore a code for 

\smallskip
\textsf{\begin{tabular}{p{10cm}} Activate $d[r]$ whenever the proposed code $r$ fits at least $k_d$ steps of recent memory \end{tabular}}

\smallskip
should itself fit the sequence $c_1, \dots, c_n,$ so that under the premise of being simple enough, it could be established by diagonalization.

\smallskip
{\bf 4.} If we assume that a self-editing code is equipped with an internal representation of the environment, the above mental experiments suggest that it is possible to predict by diagonalization (in a similar way to that  we actually do) smooth environmental  changes. An obvious example of this, regards the concept of velocity: Mental experiment 2 in this case, can be used as a basis to establish that acceleration may also be detectable.

\subsection{Other forms of diagonalization.} 

\subsubsection{Statistical form.}
Let us consider again the case, in which by seeing a surviving or successful sequence $c_1, \dots, c_n$ and without knowing nothing about its environment, we wish to help it by remarking patterns that are probably necessary to be perpetuated. The first kind of diagonalization procedure we wish to introduce here, has to do with observing that a simple algorithmic instruction $R,$ may fit a  percentage (and not all) of the transitions $c_i \mapsto c_{i+1},$ for $i < n.$ In this case, and in a need to help the self-editing code $c_n,$ we could reprogram or advise it to use $R$ with the same relative frequency as we see it in the all sequence. This advice may concern either the calculation of its descendants, in the case of proliferating transitions, or the probability of using a certain differentiating code $r$ (which should be a code for $R$), in the case of non-proliferating ones. Say for example that we notice that in the sequence $c_1, \dots, c_n,$ specific sub-codes are copied in half of the transitions $c_i \mapsto c_{i+1},$ $i < n.$ Assuming that this remark is not seemingly related to a noticeable condition that is apparent in the rest of the code, we may deduce that it is advisable that these specific sub-codes should be copied in half of the descendants that are computed, or with probability $1/2.$ This statistical behavioral strategy, gives us by means of the complete memory version of the basic self-editing principle, combined with the memory lemma, a statistical version of diagonalization.  

Indeed, given a searcher $\Delta_s,$ we may alter the functioning of the tester, to check the relative frequency that a given proposed code $r,$ fits the transitions $c_i \mapsto c_{i+1},$ $i< n,$ of a given sequence $c_1, \dots, c_n,$ and take the decision to use $r$ with the same relative frequency as a differentiating code. If $\delta$ is a code of the overall procedure and is activated in $c_n = c_n[\overline{\delta}],$ then by the self-editing principle and the memory lemma, $c_n$ can perform the same algorithm and take the same decision. One should notice here that as in the case of the simple form of diagonalization, decisions can be made either to use $r$ as a permanent or temporary differentiating code, or as a $\phi$-differentiating code for some particular address $\phi.$ Again, the length of the recent memory that $r$ fits with a given relative frequency, plays an important role to establish it as a permanent or temporary possible behavior. The most crucial examples of the application of this procedure, are for the first part the establishment of a variety of differentiating codes that are used to compute descendants in a proliferating step, yet also the probability that is used for a non-proliferating self-editing code,  to calculate its next state. This last has also to do with the priorities of the proposed by the searcher code. The case will be examined in more detail, later on. 

It is worth to notice here, that the statistical version of diagonalization described above, includes the simple form that we have already seen. Moreover, it is clearly more useful in a practical way. The interesting fact that holds also true, is that the statistical form (as we are going to see) can be established by the simple form. Roughly speaking, a self-editing code using the simple form, may notice in its history that it successfully establishes the use of differentiating codes, with the same relative frequency that are used in its history. Of course such an observation to be valid, requires the choice of a surviving sequence of computations (which can be performed by the environment) or of a successful sub-sequence, which should have been selected by the self-editing code itself, as we are going to see below.

\subsubsection{Parallel diagonalization.} Let us return once again to consider the case in which we would like to program or advice a self-editing code $c = c_n,$ by trying to find patterns in its surviving history $c_1, \dots, c_n.$ Let us for the moment assume, that to every code there correspond two addresses $\theta_i$ and $\theta_o$ in which environmental input and environmental output are registered respectively. Suppose also that the demand of the environment is a simple enough relation $R,$ (so that it can be observed by us) between the input and the output of a code and that this relation is time independent, i.e. it remains the same during our experiment. Due to the assumption that $c_1, \dots, c_n$ is a surviving sequence, we deduce that for every $k<n,$ the content of $c_k$ in its input $\theta_i$ is related with the content of $\theta_o$ in $c_{k+1}$ with this same relation $R.$  I.e. denoting by $c \restriction \theta$ the value of the content of $c$ in address $\theta$ we should have that 
\[ R(c_k \restriction \theta_i) = c_{k+1}\restriction \theta_o, \]
for every $k < n.$ Thus, it is evident that if $r$ is a code for the algorithm described as:

\smallskip
\textsf{\begin{tabular}{p{10cm}} Output to the environment the code that results by computing $R(i),$ where $i$ is the code of the environmental input, \end{tabular}}

\smallskip
fits the sequence $c_1, \dots, c_n,$ so that it can be guessed as a strategy in case that $c_n$ uses diagonalization. The problem we would like to address here, is that the structure of the sequence $c_1, \dots, c_n$ does not indicate the nature of the problem. It should be much more natural to ask to find a code $r$ that fits the set of transitions
\[ \{ c_k \mapsto c_{k+1}: k < n\} = \{c_1 \mapsto c_2, c_2 \mapsto c_3, \dots, c_{n-1} \mapsto c_n\},\] i.e. to find a code $r$ such that for all $k<n,$ $\alg(r)(c_k) \sqsubseteq c_{k+1}.$ Of course, in the particular example, the testing algorithm is equivalent to the one that results from the sequence $c_1, \dots, c_n,$ although this is not true for all cases. We will call a diagonalization of the above form, i.e. one that uses a set rather than a sequence in the place of the testing algorithm, a {\em parallel diagonalization}. Intuitively, whereas sequential diagonalization is useful both to detect evolutionary steps of the same  kind in an evolving self-editing code or on the other hand a smoothly changing environment, where the order of a sequence plays an important role, parallel diagonalization detects a relation on a set of transitions. 

An important example of parallel diagonalization is {\em free-floating DNA} that is able to communicate genes involved in antibiotic resistance, see for example \cite{calderon-franco}.

Communication between humans is a similar example.

Notice that in both examples above, there is actually no sequence to consider. Moreover the code generator is non-existent and has been replaced by one or more atoms of a certain population (the ones that transfer the particular message).  

Let us now consider the case of a set $\{s_1, \dots, s_n\}$ of sub-codes of a given self-editing code $c.$ It is clear that parallel diagonalization over the development of the set $\{s_1, \dots, s_n\}$ can create common features of the sub-codes $s_1, \dots, s_n.$ That is, permanent decisions that concern all of the codes in common can be considered as a common extension of all of the codes. This observation can be used to explain a lot of phenomena of abstraction. To see this, assume moreover that the self-editing code is equipped with an  internal representation of the environment. In this case, parallel diagonalization over the environmental objects $s_1, \dots, s_n$ can reveal their common nature as an abstraction. If for example $s_1, \dots, s_n$ are specific instances of dogs, created by the experience of interacting with actual corresponding dogs, then parallel diagonalization over this set, can potentially create the abstract notion of a dog. Similarly, in the case that $s_1, \dots, s_n$ are couples of various objects, a parallel diagonalization can potentially create the abstract notion of the number $2,$ and so on.

\section{Surviving sequences and successful sub-sequences.} \label{subsequenceDiagonalization} In the aforementioned mental experiments, let us consider the case where the answers for the not intended to be guessed digits, are given by means of proliferating steps in which at least one descendant gives the correct answer. 

 During the first mental experiment, since we have considered the first three digits $0,$ $1$ and $2$ as not intended to be guessed, we may assume by the basic guessing convention that their correct guessing, results by mere chance on some of the immediate descendants of the previous code. That is, we assume that we begin with a code $c_0 = c_0[(\theta_e) 0]$ that guesses the correct number $0$ to be output just by chance, and we 'll also assume that there are $c_1$ in the immediate successors of $c_0$ and $c_2$ in the immediate successors of $c_1$ such that again by chance guess the correct number to output. That is, $c_1 = c_1[(\theta_e) 1],$ and $c_2 = c_2[(\theta_e) 2].$ It is clear now, that diagonalization should instruct $c_2$ to follow the decision to compute all its immediate descendants by using $\textsf{add}(1)$ as a $\theta_e$-differentiating code, therefore computing all its immediate descendants so as to give the correct answer. In such a case this should be regarded as a successful outcome to our mental experiment.
 
 The situation becomes a bit more complicated, assuming that such environmental demand from the code, as to output the sequence $0, 1, 2, \dots$ is only local. (Notice that this actually happens in the second mental experiment). This is certainly so, since at the moment that this demand stops to incur, the decision to use $\textsf{add}(1)$ as a $\theta_e$-differentiating code for all descendants would immediately result in the surviving of none of them. (All of them would reply wrongly to the environment). Thus in this case the understanding of the local nature of such a requirement should result in a more conservatory decision of applying the corresponding differentiating code, as for example:
 
 \smallskip
  \textsf{ \begin{tabular}{p{10cm}}
Augment the percentage of immediate descendants that are computed by using $\textsf{add}(1)$ as a $\theta_e$-differentiating code. \end{tabular}  
} 

\smallskip
Notice here that the exact augmentation of such an instruction, being a parameter of the system, may also be the result of diagonalization as suggested by remark 2 above.

Such conservatory behavior is certainly in this case, a much better strategy when it comes to a surviving related problem. Clearly, a local demand can be distinguished from a non-local one, by the length of the time past that exists and therefore by the length of the recent part of the surviving sequence that is forced by selection to be in accordance with the demand. On the other hand, since this conservative strategy is a better one, assuming that in a surviving sequence all have been done well by chance, it would be expected for this strategy to be kept, exactly in the cases where the chosen differentiating code fits a short part of the sequence. Therefore a code for 

\smallskip
\textsf{\begin{tabular}{p{10cm}} If the chosen differentiating code fits a short part of the recent sequence in memory, then apply it conservatively, \end{tabular}}

\smallskip
should be a code that fits the entire surviving sequence, so that it can be established by diagonalization as well. 

This conservative behavior which is related to temporal aspects of the environment, may as well explain the problem of {\em variation of species.} (See section \ref{evidence} and \cite{livnat} for more details on the subject). 

On the other hand, a differentiating code that fits the entire surviving sequence, as the preceding one, should be regarded as a better candidate to be applied permanently and in all (or almost all) descendants. Again, as one can easily see, this should be apparent as well in the history of a code. 

\label{stabilization} An interesting example of this long-term memory fitting, is the case where the history of the surviving predecessors of a code $c$ for a long period indicates that they remain the same, i.e. there is a sufficiently large positive integer $n$ such that the most recent $n$ terms of its history $c_1, c_2, \dots, c_n$ are the same as $c.$ (That is for every $i \le n,$ $c_i = c$). In this case, assuming that $c_n = c$ decides its successors using diagonalization, it should be expected that its decision would be to copy itself to (probably almost) everyone of them. 

Clearly this remark holds also true assuming that parts (instead of all) of the code remain the same for an extended period in its memory. This function can clearly lead to a stable program of a given self-editing code, in the sense that it could learn this way to inherit useful parts (judged by selection itself) to all of its descendants.

Finally, let us observe here, that the decision of a long-term memory fitting of a differentiating code is not necessary to use a really long-term memory, since a code may delete its memory and at the same time store the information that a given differentiating code fits the memory so far, as a simple evaluation regarding the differentiating code.

Notice that the previous observations apply equally well to a successful sub-sequence of a non-proliferating self-editing computation. In fact, it is interesting to observe the relation between long-term memory fitting of a differentiating code to the difficulty that is presented by human beings to alter a prolonged habit.

\subsection{Diagonalization over successful sub-sequences.}  The problem we would like to address here, is whether it can be learned by a self-editing code, to apply diagonalization to successful sub-sequences. Of course, the candidate method for learning such a process would be diagonalization itself.

Let us consider  $\delta'=\delta[(+\theta_t) x_1, \dots, x_m]$ to be a code for a diagonalizing algorithm with $x_1, \dots, x_m$ being the testing sequence. I.e., $\alg(\delta')$ is designed so as to find the simpler code $r$ that fits the sequence $x_1, \dots, x_m.$

Notice now, that if $c_1, \dots, c_n = c_n[(\theta)\delta']$ is the history of the code $c_n$ and $k\le n,$ then the function
\[ c_1, \dots, c_k, \dots, c_n = c_n[(\theta^\frown \theta_t) x_1, \dots, x_m] \mapsto c_n[(\theta^\frown \theta_t) x_1, \dots, x_m, c_k] \] 
that appends $c_k$ in address $\theta^\frown \theta_t$ of $c_n$ where the testing sequence (i.e. the sequence to be diagonalized) is situated, is clearly algorithmic, so that by the basic self-editing principle, it can be triggered from within $c_n.$ Therefore a self-editing code may decide about the construction of the testing sequence over which diagonalization occurs.

\smallskip
Assume now that 
\[ c_1\mapsto c_1^1 \mapsto \cdots \mapsto c_1^{n_1}\]
is a self-editing computation without proliferating steps. Let us moreover assume that $c_1^{n_1}$ undertakes a proliferating step by giving as immediate descendants variations of $c_1$ and let $c_2$ be one of them. As before, let
\[ c_2 \mapsto c_2^1 \mapsto \cdots \mapsto c_2^{n_2}\]
be the self-editing computation that starts with $c_2$ without proliferating steps. Continuing in this manner, we may come to a surviving sequence $c_1, c_2, \dots, c_k,$ where for every $i\le k,$ $c_i$ produces a self-editing computation 
\[ c_i \mapsto c_i^1 \mapsto \cdots \mapsto c_i^{n_i}.\]
We may call the sequence
\begin{equation} \label{composite} (c_1, \dots, c_1^{n_1}), (c_2,  \dots,  c_2^{n_2}), \dots, (c_k, \dots, c_k^{n_k})\end{equation}
 a sequence of cycles or a {\em composite sequence}. The important assumptions we make are 

\begin{enumerate} \item[(i)] For every $i \le k,$ the computation $c_i, c_i^1, \dots, c_i^{n_i}$ has no proliferating steps.
	\item[(ii)] For every $i < k,$ the step $c_i^{n_i} \mapsto c_{i+1}$ is a cycle, i.e. $c_{i+1}$ is computed by $c_i^{n_i}$ as a variation of $c_i$ instead of itself.
	\item[(iii)] For every $i<k,$ the self-editing step that $c_i^{n_i}$ undertakes is a proliferating one, thus $c_1, c_2, \dots, c_k$ is a surviving sequence.	
	\end{enumerate}
Later on, we are going to omit assumption (iii), and replace it with the demand that $c_1, c_2, \dots, c_k$ is instead {\em successful}. The idea is that composite sequences serve as a kind of a general scheme to be used for recursion. On the basis of this scheme lies survivability. What we intend to explore here is that survivability may serve on the basis of this recursive hierarchy to characterize success. 

Composite sequences are very useful towards the purpose of finding out the necessary ingredients of a successful computation. Indeed, since $c_1, \dots, c_k$ is a surviving sequence, one can deduce that computations made in between $c_i$ and $c_{i+1},$ $i < k,$ that is, every computation \[c_i, c_i^1, \dots, c_i^{n_i}\] should be successful. So, again, any pattern recognized in these computations should be perpetuated.  

In a composite sequence, such as \eqref{composite}, we may distinguish the self-editing steps in two categories: The {\em certain} ones which is thought to be all the steps $c_i^{n_i} \mapsto c_{i+1},$ for all $i < k.$

The {\em uncertain} steps which are the steps $c_i^m \mapsto c_i^{m+1}$ for all $i \le k$ and $m < n_i.$

The terms {\em certain} and {\em uncertain} are used here to express exactly that any cycle in \eqref{composite} is a product of selection while any uncertain step is not.

Since now the sequence $c_1, \dots, c_k$ is a product of selection, we may assume that any sequential diagonalization that takes place during the self-editing computation of a certain step uses as a test sequence the all history of the code. 

The conventions we make about uncertain steps, are that
failed uncertain self-editing steps are selection forgiven, yet failed diagonalization during an uncertain step is not. This consideration roughly mirrors the belief that while false individual answers to the environment, should be thought of as necessary in a learning procedure, false conclusions about the environment (which result from a failed sequential diagonalization) may lead to massive failed answers and therefore should be expected to be condemned by the environment. 

Because of this consideration, since the $i$-th cycle $c_i, \dots c_i^{n_i}$ is expected to contain failed answers, which are not supposed to be contained in a sequential diagonalization process, the testing sequence of such a diagonalization during an uncertain step $c_i^m \mapsto c_i^{m+1}$ should not contain the sequence $c_i, \dots, c_i^m$ itself but rather the sub-sequence of it that contains the successful answers.

Let us now take for granted that there exists a testing algorithmic procedure $T$ (i.e. one that outputs \textsf{true} or \textsf{false}), such that $T(c) = \textsf{true},$ if and only if $c$ should be included in the testing sequence of a diagonalization during an uncertain step. Given the possibility of choosing the correct codes to be included in this testing sequence, it is likely that this should have happened in a surviving sequence of codes as $c_1, \dots, c_k.$ Therefore, it should be expected that if  $r$ is a code for 

\smallskip
\textsf{Let $x$ be the input.\\
	If $T(x) = \textsf{true},$ then append $x$ to the testing sequence,}

\smallskip
should fit (at least statistically) the sequence \eqref{composite}, that is, if for some $i\le k,$ and $j< n_i,$ $T(c_i^j) = \textsf{true},$ then $c_i^j$ (which is in fact the input of $\alg(c_i^j)$) should be included in the testing sequence of $c_i^{j+1}.$ As explained above, this is so, assuming that successful diagonalization should lead to a code with greater ability to survive. Therefore, the use of $r$ may be established by diagonalization that occurs in a certain step. Notice that this is also a concrete example of the ability of diagonalization to evolve itself in the frame of self-editing.

\smallskip
At this point, we can make a step further, by  replacing the surviving sequence $c_1, \dots, c_k$ in \eqref{composite}, with  a successful one. Assume for example that for $i \le k,$  $c_i, c_i^1, \dots, c_i^{n_i}$ is a computation of the diagonalizing unit, resulting in a successful answer to the environment which is registered exactly at the environmental output of $c_i^{n_i}.$ In this case, diagonalization over the entire sequence \eqref{composite} may lead to a better functioning of diagonalization itself. Generally speaking, diagonalization to composite sequences such as \eqref{composite} is in accordance to whatever we experience as a procedure of learning. Indeed, such a procedure should have as necessary components the notions of success and failure as well as repetitive tasks. Let us see some concrete examples of this procedure in action:

\smallskip
{\bf Example 1.} Consider the 1st mental experiment, where a self-editing code has to guess the dots in the sequence
\[ 0, 1, 2, \dots.\]
In this case, we will assume that in non intended to be guessed digits, the algorithm takes guesses in chance, just as we would do during the same probation. We will make use of the basic assumption here, in the form that we 'll take as granted that the non intended to be guessed digits, are indeed eventually found. Assuming an input address, let us call it envronmental input, and a 0-1 feedback in this same address that encodes for failure or success of the answer respectively (that is, 0 for a failed answer and 1 for a successful one), we may replace the testing algorithmic procedure $T$ such that $T(c)$ is \textsf{true} if and only if 1 is registered in the environmental input of $c.$ By the discussion above, we may begin with a code $c_0$ with a stable diagonalizing functioning, so as to append a code $c'$ from its history to the sequence to be diagonalized in all cases that 1 is registered to the environmental input of $c'.$ Assuming that $c_i, c_i^1, \dots, c_i^{n_i}$ are the attempts to guess the $i$th digit, by the basic assumption, we may ensure that $c_i^{n_i} = c_i^{n_i}[(\theta_e) i],$ that is $c_i^{n_i}$ has managed after some attempts to guess the correct digit (which is exactly $i$ in this case). Therefore diagonalization now should take place over the sequence $c_0^{n_0}, c_1^{n_1}, c_2^{n_2},$ since these are the codes that take the feedback 1 in their environmental input. This is indeed the correct sequence for diagonalization to infer that the code must proceed adding 1 in the address of environmental output, guessing thus successfully the rest of the digits.

It is easy to see that this same line of thoughts would lead in success in the second mental experiment as well.

\section{Diagonal generalizations.} At this point, we are interested to study the application of diagonalization for the establishment of a code with a variable sub-code. A very often occurring such example, concerns the potential ability to apply as a decision code, a code of the form $s[r]$ whenever $r$ meets specific requirements. We assume that these requirements consist of a testing algorithm $T$ (i.e. one that outputs only \textsf{true} or \textsf{false}).

Consider the following composite sequence as an either surviving or successful one:
\begin{equation} \label{composite1} (c_1, c_1^1, \dots, c_1^{n_1}), (c_2, c_2^2, \dots, c_2^{n_2}), \dots, (c_k, c_k^1, \dots, c_k^{n_k}) \end{equation}

Let us assume a code $t = t[ \varnothing]$ for $T$ and a code $s = s[\varnothing],$ for the decision to be taken. In this case, we regard the empty addresses in $t$ and $s$ respectively, to be exactly the addresses to be filled with the variable code $r.$  Let us assume further that $\alg(t)$ may depend on a second argument as well, being either an external condition or an internal one. (The case that this doesn't happen is a trivial one, in which either $t[r]$ is always true or always false). Since the general case we consider is that this condition is encoded as a sub-code of the self-editing code itself, either as an environmental input in its memory (in the case of the condition being external) or not (in the case that the condition is internal), we may assume that $t$ accepts the self-editing code in a second address, that is we assume that $t$ is of the form $t[ \varnothing, \varnothing].$ Under this terminology, $T(r)$ is true in an instant of \eqref{composite1} that is determined by an $i \le k$ and a $j < n_i,$ if and only if $\alg(t[ r, c_i^j]) = \textsf{true}.$

Our basic assumption is that the application of the algorithmic instruction (let us call it $M$):

\smallskip
\textsf{If for some $r,$ $\alg(t)(r, c)$ is true, then take the decision $\alg(s)(r),$ for this same $r$}

\smallskip
is a necessary condition for either surviving or succeeding. Therefore, since \eqref{composite1} is indeed either surviving or succeeding respectively, we deduce that whenever for some $i\le k$ and $j < n_i,$ $\alg(t)(r, c_i^j)$ is true, then $\alg(s)(r)$ fits (obviously by chance up to now), the computational step $c_i^j \mapsto c_i^{j+1}.$ Thus, if $m$ is a code for $M,$ we infer that $m$ fits $c_i^j \mapsto c_i^{j+1}$ for all $i \le k$ and $j< n_i.$ Consequently, by diagonalization, the application of $m$ may be established to every computational step thereafter.

\smallskip
{\bf Remarks.} 1. The above analysis has been done on the premise that $m$ is an adequately simple code related to the sequence \eqref{composite1}. Generally, even a not so simple code, may be found to be the simpler one fitting a long and varied experience. This last observation is adequately backed up by Probability theory. On the other hand, the simpleness of $m$ relies on the simpleness of $t$ and $s.$ Regarding this point of view, diagonalization can be used as well in this case to render as simple, codes that have been proved useful by experience. (For a simple case of this procedure, see Example 2 below).

2. Obviously, the same reasoning can be used to establish diagonal generalizations for more than one code-variables, (in addition to $r$).

\smallskip
{\bf Example 2.} \label{searcherTraining} Up to now we have assumed that the priorities of the set of proposed codes in a diagonalization process is fixed. However, if $\delta_s$ is the searcher's code, there are a lot of algorithmic ways to change the priority of a given code $r,$ the simplest one probably being to add a corresponding instruction $s[r]$ to $\delta_s$:
\[ B(r) : \delta_s \mapsto \delta_s[(+) s[r]], \]
where $s[r]$ is a code for 

\smallskip
\textsf{Output $r$ with priority $n.$}

\smallskip
More complex (yet also more interesting) such algorithmic changes of the priority of a code to a searcher, occur, assuming that codes are constructed by the searcher by composing elementary instructions and have to do in this case with altering conditional probabilities of the composition process inside $\delta_s$ itself. We are not going for the moment to deal with such complex situations, yet generally if $B(r)$ is such an algorithm for altering the priority of a code $r$ in a searcher with code $\delta_s,$ since $\delta_s$ should be a sub-code of a self-editing code that uses diagonalization, it follows easily from the basic self-editing principle, that a self-editing code may undertake the decision to alter the priority of proposing a differentiating code. 

Let now $x \mapsto x'$ be a self-editing step in which $x$ has decided randomly to raise the priority of $r$ as a proposed by the searcher code in the diagonalization process. Assume moreover that this decision is successful. If
\[ \{(x_1, x_1^1, \dots, x_1^{n_1}), (x_2, x_2^1, \dots, x_2^{n_2}), \dots, (x_k, x_k^1, \dots, x_k^{n_k})\}\]
is the set of the computations of the diagonalization unit in the history of $x,$ it is expected that the relative frequency of these cases that $r$ has been accepted by the testing sequence as a fitting code, exceeds the relative priority of $r$ being proposed by the searcher. This should be correct, since raising the priority of $r$ as a proposed code has been considered successful. 

Therefore if $c_i \mapsto c'_i,$ $i =1, \dots, m$ are successful self-editing steps in the history of a self-editing code $c,$ in which the priorities of the proposed codes $r_i,$ $i=1, \dots, m$ are respectively raised, then for every $i=1, \dots, m$ the step $c_i \mapsto c_i'$ falls in the category of the previous observation. Therefore a code for the algorithm

\smallskip
\textsf{\begin{tabular}{p{10cm}} If the relative frequency of the acceptance of any code $r$ during the diagonalization processes, is greater than its relative priority to be proposed, then raise this same priority \end{tabular}}

\smallskip
should fit an either surviving or successful sub-sequence of the history of $c.$ Hence, such an instruction can be established. Therefore a self-editing code can learn by diagonalization to render as simple often useful proposed codes. Notice that this is also a concrete example of how diagonalization can be used in the frame of self-editing to improve itself.    

 Notice that the stratagem described above as well as the statistical form of diagonalization is roughly reminiscent of the {\em multiplicative weights update algorithm} (MWUA), an efficient optimization algorithm that has been discovered many times in computer science, statistics, and economics \cite{barton}.

\smallskip 
{\bf Example 3.} \label{undecidability} The question whether a given code fits a sequence is in fact a non decidable problem, since it has been proved by A. Turing in his famous {\em Halting problem} \cite{boolos}, that given two codes $r$ and $c,$ there is no algorithmic way to decide whether $\alg(r)(c)$ eventually stops and gives an answer. (It may run for ever as well). In spite that fact, in our every day lives, we may conclude that $\alg(r)(c)$ is not going to stop, by simply waiting enough time to do so. Similar decisions are possible for a self-editing code that uses diagonalization: Let
\[ (c_1, c_1^1, \dots, c_{n_1}), (c_2, c_2^1, \dots, c_2^{n_2}), \dots, (c_k, c_k^1, \dots, c_k^{n_k})\]
be a composite sequence such that $c_1, \dots, c_k$ is either a surviving or a successful sub-sequence. Assuming now that for every $i = 1, 2, \dots, k,$ $c_i, c_i^1, \dots, c_i^{n_i}$ contains the calculations of the tester to see whether a given proposed code fits or not the testing sequence, we can deduce that there should be an upper limit of calculation steps (and therefore of time) that are performed in every such calculation. This is indeed so, since above this limit $c_1, \dots, c_k$ wouldn't be surviving or  successful respectively. Thus a code for 

\smallskip
\textsf{\begin{tabular}{p{10cm}} For every proposed code $r$ and every code $c$ in memory, wait at most $n$ steps to calculate $\alg(r)(c),$ \end{tabular}}

\smallskip
should fit the sequence and therefore could be established in the process of diagonalization.

\smallskip
{\bf Remark.} It is interesting to notice here that assuming an internal representation of the environment, the same procedure as above can be used to establish non-decidable truths by experience, (that is by diagonalization). Notice also that the procedure to establish such truths by experience, is very common in our functioning as human beings. Of course such a method of establishing truths does not contain certainty and for this reason we also use logical deduction, something that is much more complicated in nature to examine here. One should bear in mind though, that logical deductions require axioms and axioms are not subject of proofs, but solely of experience.  

\smallskip
{\bf Example 4.} \label{homeostasis} {\em Homeostasis}. It is evident that no parameter in the evolution of a self-editing code, can be successfully being repeatedly increased to infinity. Even  if there is some possible benefit of doing so, this should be at some point counteracted by the cost of retaining a sufficiently large number for this parameter, in whatever form it is. Thus, a surviving or successful sequence of self-editing codes, should have found at least locally optimal values to a lot of parameters. Considering a sequence 
\[ (c_1, c_1^1, \dots, c_1^{n_1}), (c_2, c_2^1, \dots, c_2^{n_2}), \dots, (c_k, c_k^1, \dots, c_k^{n_k}),\]
of such successful outcomes, diagonalization upon the sequence could provide strategies for achieving such an optimal value. In the case of a surviving sequence, conservative strategies (for example to generate a sole descendant of the same value at each step), combined with diagonalization, (since in the case of a locally optimal value, this same value should remain constant in recent memory), provide such a strategy. Such a conservative strategy can be also used in a successful sequence, where the role of selection in this case may be played by either internal or external reward. In fact we are aware of such a strategy, in the cases where emphasizing gradually a particular characteristic of our personality, can reach a point that this becomes inappropriate. The usual reaction to such a feeling, is simply to stop emphasizing it more.

\section{Learning specialization.}
 \label{learningSpecialization}
Let us consider the problem of finding a code $r$ that fits a particular test $T = \alg(t),$ in order to use it in a decision with code $d[\varnothing].$ (Notice that the above consideration describes in an abstract form the problem of applying diagonalization). Assume now that $c$ is a self-editing code that uses a decision algorithm with code $\delta[t'] = [\delta_s, \delta_t[t']]$ where $t'$ is the code of its testing algorithm. Assume moreover that the code $d[\varnothing]$ of the decision to be made lies in some address of $c= c[d[(\theta) \varnothing]]$ so that $\theta$ is the (absolute relatively to $c$) address of the argument of $d[\varnothing]$ and let $\phi$ be an available address of $c.$ The situation can be summarized by denoting 
\[ c = c[\delta[t'], (\theta) \varnothing, (+\phi) \varnothing].\]
 Using $\phi$ as a $\theta$-differentiating address, one can use the decision code $\delta[t]= [\delta_s, \delta_t[t]],$ (notice that we have replaced the appearance of $t'$ with $t$), as a $\theta$-differentiating code by simply copying it to $\phi.$ The idea here is that as we saw in Example 2, page \pageref{searcherTraining}, the searcher code $\delta_s$ is expected to be favorably evolved and thus in this case should be able to be specialized as a searcher of $\theta$-differentiating codes. Given the code $t,$ the above process can be described as the function
\[ c[\delta[t'], (\theta) \varnothing, (+\phi)\varnothing] \mapsto c[\delta[t'], (\theta)\varnothing, (+\phi) \delta[t]],\]
which is clearly algorithmic, so that by the basic self-editing principle, the decision of using it can be made by $\alg(c)$ itself. Hereafter we are going to abbreviate such a decision as 

\smallskip
\textsf{Find a code $r$ such that ...}

\smallskip
and we will refer to it as  a {\em $\theta$-specializing instruction}. 

\smallskip
Given now an address $\theta$ in an either surviving or successful sequence $c_1, \dots, c_m$ of self-editing codes, and assuming that $\delta[t']$ is a code of a diagonalizing algorithm, it is possible that accepted by the tester proposed codes that are used as $\theta$-differentiating in $c_1, \dots, c_m,$ may be in disagreement with the initial priorities of the proposed by the searcher $\delta_s$ codes. This could be proved a very plausible condition that should trigger a $\theta$-specializing instruction. For the remaining of this remark, let us refer to this condition as the {\em rate disagreement condition for $\theta$}.

Assuming that the information of the priorities of the proposed codes can be detectable in the code of the searcher $\delta_s$ and since the rates of the acceptance of the proposed codes can be detected in the calculation $c_i, c_i^1, \dots, c_i^{n_i}$ of the diagonalizing procedure for every $i \le m,$ a rate disagreement condition for $\theta$, may be checked in the composite sequence
\begin{equation} \label{composite3} (c_1, c_1^1, \dots, c_1^{n_1}), (c_2, c_2^1, \dots, c_2^{n_2}), \dots, (c_m, c_m^1, \dots, c_m^{n_m}),\end{equation}
so that a $\theta$-specializing instruction can be triggered when necessary by a self-editing code that uses rate disagreement conditions to trigger $\theta$-specializing instructions. One should notice that assuming enough experience, diagonalization can suggest such a behavior:

Indeed, assume that $c_1, c_2, \dots, c_m$ is an either surviving or successful sequence of a self-editing computation where specializing instructions have been decided randomly. Since in this particular sequence, these instructions would be successfully taken and the rate disagreement condition can be checked by a self-editing algorithm, it follows that a code for 

\smallskip
\textsf{\begin{tabular}{p{10cm}} If the rate disagreement condition holds for an address $\theta,$ then decide a $\theta$-specialization, \end{tabular}}

\smallskip
should fit the sequence, so that a self-editing algorithm may establish it and thus learn to use it by diagonalization.

\smallskip
{\bf Remarks.} {\bf 1.} It should be interesting to notice here, that specialization may as well occur in sub-addresses of the searcher $\delta_s$ of a decision unit. The interest of this observation lies onto the fact that repeated such specializations can result in an hierarchy of decision units. The formation of such an hierarchy may follow the diagonalization principle as well, something that as we will see in subsection \ref{InternalRepresentation} of page \pageref{InternalRepresentation} can be used of as a way to accomplish abstractions about the environment.

\smallskip
{\bf 2.} In the case of a $\theta$-specializing instruction of a diagonalization unit, as we have seen, the testing algorithm of the tester, may be a subject of diagonalization, so that it can be formed by using it. Nevertheless, it is obvious that the testing algorithm in this case, should be replaced by the testing algorithm that checks the fitting of the proposed codes to sequences of successful values of $\theta$.  

\smallskip
{\bf 3.} It is evident that the statistical form of diagonalization generalizes the simple form. On the other hand it may be deduced as a strategy by the simple form as follows:

Let us assume a self-editing code which in its history has fixed successfully the weights of various  differentiating codes that it uses. For any such fixed weight $w$ of a differentiating code $r,$ it is expected that $w$ should match the relative frequency that $r$ fits the sequence in its history, so that a code for 

\smallskip
\textsf{\begin{tabular}{p{10cm}} Find codes $r$ that fit the history sequence with a positive relative frequency $w$ and use them as differentiating codes with the same as $w$ weight, \end{tabular}}

\smallskip 
should fit an either surviving sequence or successful subsequence of its history. Notice now that this same code performs essentially diagonalization in its statistical form.

\section{Hints for further research.} \label{hints}

\subsection{Language.} Up to now, we have used combined sequences of the form
\[(c_1, c_1^1, \dots, c_1^{n_1}), (c_2, c_2^1, \dots, c_2^{n_2}), \dots, (c_k, c_k^1, \dots, c_k^{n_k}) \]
in order to assign values of success and failure in steps in internal cycles, assuming that there is such an assignment to the external ones either by the environment itself or by recursion.

However, one can use combined sequences as well, to assign success and failure in longer time scope, based in the assumption that such notions (success and failure) have been assigned in shorter time scope. Naturally, shorter time scope is represented here by the internal cycles and long time scope by the external ones. This is reminiscent of two situations which are not completely irrelevant between them.

\smallskip
The first one, is the way we learn in a completely new environment. This is particularly evident in the case of children whose memory is of limited time scope and lengthens with the years of age. (See also \cite{analogy}). On the other hand, what kind of an algorithm could regulate its memory according to its knowledge if not a self-editing one. The same reasoning can provide a plausible cause for the following phenomenon: It is a common knowledge, as noticed also in \cite{analogy}, that the subjective perception of elapsed time shortens with age. In the frame of the theory that is presented here, a plausible explanation would be that in the course of time less self evolution occurs, as the later approaches its upper limit, and on the other hand it occurs a better understanding of the environment, thus less environmental facts to be related to the self and between them.

\smallskip
The second one is the way that children learn to speak: They begin using syllables (notice that as an education practice we tend to reward them for the correct ones), then try to combine them in words, phrases, etc. The importance of induction in language learning has been also emphasized by  Hauser, Chomsky and Fitch in \cite{hauser}.

\subsection{Structure.} As stated in \cite{clune2}, a key driver of the capability of biological organisms to quickly adapt to new environments, is the widespread modularity of biological networks, that is, their organization as functional, sparsely connected subunits. Despite its importance and decades of research, there is no agreement on why modularity evolves \cite{wagner, wagner2, espinosa-soto, kashtan, kashtan2, clune2}.

On the other hand, assemblies of neurons \cite{papadimitriou, hebb, abeles, harris, buzsaki, miller, buzsaki2, eichenbaum, ranhel}, are large populations of neurons believed to imprint memories, concepts, words, and other cognitive information. Their formation is not yet understood \cite{buzsaki}.

In both cases, our proposal is that structure emerges in order to facilitate diagonalization in particular features of the overall program. Notice that this view is in accordance to the theory that modularity arises in biology to enable modifying one subcomponent without affecting others \cite{espinosa-soto} and also with the hypothesis that modularity (in biology) emerges because of rapidly changing environments that have common subproblems, but different overall problems \cite{kashtan, kashtan2}, since the structure that induces parallel diagonalization is exactly the same. The general problem of formation of the structure seems to be relative with the problem of language, since the structure of a self-editing code defines a language to talk with itself. However, we have already seen definitions of modules by a self-editing code, for example the formation of a differentiating code. Actually, in Mental experiment 2 of page \pageref{mentalexperiment2}, we essentially used the formation of a differentiating code in order to diagonalize upon it. 

The general guess we can make here about the formation of units of structure, is that sequential diagonalization can detect codes that are usually computed in a sequence of computations, so that a self-editing code can store them for future use, saving thus computing power. The strategy itself of doing so, can also be detected by diagonalization, assuming that successful storing of codes has occurred a number of times before. One should notice here that using this strategy, is exactly what we do in language when we often use an object or an idea that has a complex description: We simply define it by using either a shorter phrase or even a single word. 
   
\subsection{Internal representation of the envirnonment.} \label{InternalRepresentation} It has been hypothesized that the human brain encodes for the environment and by doing so it can generate predictions and recall memories, see for example \cite{buzsaki2, miller, neisser, bar}. We would like here to describe a way that such an encoding could take place, in the case of self-editing computations.

So, assume that $c_1, \dots, c_n$ is an either surviving or successful sequence of self-editing codes and moreover that in every code $c_k,$ $k \le n,$ there are addresses for the input $i_k$ that $c_k$ receives and for the output $o_k$ that sends to the environment. Assuming diagonalization, $c_n$ can decide the probably successful output to $i_n$ (that has received), by finding a code $r$ such that for all $k \le n,$ $\alg(r)(i_k) = o_k.$ The suggested output then should be $\alg(r)(i_n).$ It is obvious on the other hand, that such a strategy as diagonalization, requires a large amount of storing memory, something that quickly renders it unpractical. This problem may have a similar to diagonalization answer: Assuming that one has in his disposition a sufficiently large amount of codes $r_1, \dots, r_m$ that are either rejected or accepted by the above diagonalization process, one can search for a test code $t$ such that $\alg(t)(r_\ell),$ $\ell \le m$ is a rejection if $r_\ell$ does not fit the set of pairs $i_k, o_k,$ $k \le n$ and an acceptance if it does fit the same set of pairs. Naturally, the searcher that proposed the test-codes like $t$ above should be a subject of evolution (or learning) so that ideally could reach a point to be successful enough for its choices. Since obviously it cannot be expected that a code $t$ that describes all the environment could be possibly calculated at once, the method described above should work better combined with the idea of creating structure, where in subcodes, less complicated testing codes have to be guessed. This point of view complies with the way we learn about the world, trying to understand its features one by one. 

In a more sophisticated functionality, one should replace the pair of rejection and acceptance, which is a 0-1 evaluation, with analogue evaluations like punishing and rewarding. Thus we come to the idea of an internal representation of the real world reward and punishment, in other words, pleasure and dissatisfaction. In this case, in order to guess the testing code $t,$ that replaces diagonalization, one should have beforehand the knowledge, not only that $c_1, \dots, c_n$ is successful, but also how much successful every transition $c_k \mapsto c_{k+1},$ $k < n$ is.

Notice that the above procedure could be carried out to any subcode of the initial self-editing code to replace diagonalization on the specific subcode. Such loops is well know to exist also in the nervous system.

Another useful remark is that the procedure could be carried out by the self-editing code itself, due to the basic self-editing principle and that it can be found by diagonalization, assuming of course that some successful guesses of testing-codes have been performed by chance. This is so, since successful testing-codes should in general agree with the testing performed by diagonalization.

Let us recall at this point that due to specialization, diagonalization units can form an hierarchy if necessary, starting from the searcher $\delta_{s1}$ that proposes codes for the relation between the environmental input and output, continuing this way with diagonalization units that decide about various subcodes of $\delta_{s1}$ and so on. Repeating the above process of replacing the testing sequence of each of these units with a simpler algorithmic test which at the basis should represent simple reactions of the environment to specific actions of the self-editing algorithm, one can arrive in an hierarchy of representations of the reality and of various aspects of it by means of sequences $t_1, t_2, \dots$ of algorithmic tests. Since the essential success of forming such an hierarchy is the simplification of the actual reality regarding aspects that are irrelevant with the decisions of the self-editing code at hand,  it is interesting to investigate the application of diagonalization to the formation of the sequence $t_1, t_2, \dots.$ Such kind of diagonalization may lead for example to the perception of a ``line" as an object with only one dimension (something that certainly does not occur in nature).

\subsection{The role of sex.} Sex is nearly universal in life \cite{livnat}: ``Species that do not exchange genes in any form or manner, called ``obligate asexuals," are extremely rare, inhabiting sparse, recent twigs of the tree of life, coming from sexual ancestral species that lost their sexuality, and heading toward eventual extinction without producing daughter species \cite{livnat2}. Evolutionary theorists have labored for about a century to find explanations for the role of sex in evolution, but all 20th century explanations are valid only under specific conditions, contradicting the prevalence of sex in nature \cite{livnat}. The problem has been called ``the queen of problems" in evolutionary biology \cite{bell}." 

Our aim at this point, is to propose a plausible reason for this dominance of sexual reproduction, based on the theory that has been developed here. In fact, this explanation is quite simple, since sexual reproduction can serve as a mixture of sequential and parallel diagonalization: A code that fits two surviving sequences is much more safe to be regarded as correct, than a code that fits just one. Notice that this point of view is able to explain equally well the reason for which in sex there is attraction of the opposition: This ensures that the validity of the proposed code can be established under different attitudes. Contrary, relatives are usually excluded from reproduction for the same reason. We essentially acknowledge this principle in the usual form that a doubtful perception or idea, becomes instantly more certain in the case that it is reassured by other(s).

It is worth noticing here that the particular male-female roles in sexual reproduction is reminiscent of the relation between a searcher and a tester: Usually there are more than one male that claim a female and there is always a test (sometimes performed by the female itself) to decide among challengers. A possible explanation for this could be that the particular relationship is the product of parallel diagonalization performed in other structures. Perhaps not surprisingly, one can see the same relationship in the pair of sperm and egg.

\section{Further evidence and related work.} \label{evidence}

\smallskip
 A long-standing open question in biology is how populations are capable of rapidly adapting to novel environments, a trait called evolvability \cite{pigliucci, kirchner}.

There is currently an extended list \cite{intev, facvar, agbehe, west-eberhard, pigliucci, wagner-altenberg, frank, chastain, barton, kauffman, kirchner, kouvaris, clune2, szathmary, adams, watson, laland, fischer, livnat, livnat2, valiant, riedl} of biology related research performed by biologists and computer scientists, that supports that in the course of evolution occur phenomena like {\em learning, generalizing} and {\em thinking by analogies}. 

\smallskip
Fischer \cite{fischer} necessitates the discovery of a mathematical law about evolution as clean and central as the second law of thermodynamics.

\smallskip
Wright \cite{wright} pointed out that the frequency of an allele in a diploid locus changes in the direction that increases the population's mean fitness.

\smallskip
 The process itself of parallel diagonalization is apparent in biology: Entire genomes with their accompanying protein synthetic systems are transferred throughout the biosphere primarily as bacteria and protists which become symbionts as they irreversibly integrate into pre-existing organisms to form more complex individuals \cite{margulis}. 
 
 From \cite{dasgupta}: ``The fly's circuit assigns similar neural activity patterns to similar input stimuli (odors), so that behaviors learned from one odor can be applied when a similar odor is experienced."

Also: Two bacterial cells can pair up and build a bridge between them through which genes are transferred \cite{stearns}.

\smallskip
 S. Kauffman \cite{kauffman} argues that spontaneous order in biological organizations may enable, guide and limit selection. Therefore, the spontaneous order in such systems implies that selection may not be the sole source of order in organisms, and that we must invent a new theory of evolution which encompasses the marriage of selection and self-organization.

\smallskip
Mutations do not seem to be completely random \cite{livnat}:  

``Many important mutations are rearrangements of small stretches as well as large swaths of DNA: duplications, deletions, insertions, inversions, among others \cite{graur}. Moreover the chance of a mutation varies from one region of the genome to another and is affected by both local and remote DNA \cite{livnat2}. For a long time it has been believed that mutations are the results of accidents such as radiation damage or replication error. But by now we have a deluge of evidence pointing to involved biological mechanisms that bring about and affect mutations \cite{livnat2}.  Nearly a quarter of all point mutations in humans happen at a C base which comes before a G after that C is chemically modified (methylated) \cite{fryxell}; methylation is known to be the result of complex enzymatic processes. DNA sequences are prone to ``jump" from one place of the genome to another, carrying other DNA sequences with them \cite{graur}. Another interesting fact is that the same machinery that effects sexual recombination is also involved in mutations, and in fact produces different types of rearrangement mutations, depending on the genetic sequences that are present \cite{graur}. Finally, different human populations undergo different kinds of mutations resulting in the same favorable effect, such as malaria resistance, suggesting that genetic differences between populations cause differences in mutation origination \cite{livnat2}...Mutations are random but it may be more productive to think of them as random in the same way that the ouputs of randomized algorithms are random. Indeed, mutations are biological processes, and as such they must be affected by the interactions between genes. This new conception of heredity is exciting, because it creates an image of evolutuion that is even more explicitly algorithmic. It also means that genes interacting in one organism can leave hereditary effects on the organims's offspring \cite{livnat2}."

 Also: The total number of steps that have been done during evolution process is far less than the steps that cellphone processors do in an hour \cite{livnat}.
 
 \smallskip
 Gene interactions are considered very important to understand the hidden aspects of evolution \cite{livnat}. It is clear that they can be explained by assuming that living organisms are in fact self-editing, since in this case, a part of their program may assume a critical role in regulating (editing) another part. In \cite{livnat}, the authors raise the problem of the Preservation of Variation: Classic data indicates that, for a large variety of plants and animals taken together, the percentage of protein-coding loci that are polymorphic (in the sense that more than one protein variant exists that appears in more than 1\% of the individuals in a population), and the percentage of such loci that are heterozygous in an individual, average around 30\% and 7\% respectively \cite{nevo}. Such a number is far greater than could be explained by traditional selection-based theories \cite{lewontin, nevo}. 
 
 The theory of self-editing codes can provide an explanation for such a variation: Based on its experience via diagonalizing, it should be clearly possible for a self-editing code to compute its immediate descendants on the basis of an uncertain environmental future. (Notice that this is exactly the approach that we adopt for the same reason: One should be able at any moment to demonstrate different characteristics, depending on the environmental change). The idea is that a self-editing code should evolve to prepare its descendants for various possibilities during proliferation. This is evident by the way we handled the Mental experiments 1 and 2, page \pageref{mentalexperiment1}. Indeed, a searcher may adapt to compute the most probably appropriate descendants, as we have naively seen in Example 2, page \pageref{searcherTraining}, where diagonalization may alter the priorities of the proposed codes and as for the necessity of doing so, it should be apparent by diagonalization in the self-memory of a surviving sequence of self-editing codes, as a struggle for variation to be retained in every proliferating step. The argument is again the same and Darwinian in its nature: Not doing so, should result in the elimination of the sequence due to poor preparation for various possible environmental developments in the future. 
 
 \smallskip
 Traditionally mathematics face the phenomenon of self-reference as the exception and not the rule. Mathematical functions usually act on an argument which is usually something that does not contain the function itself.
 
 On the other hand, in the case that there exist a sufficiently large category of chemical compounds (such as DNA sequences) that are able to code for computable functions on some input and assuming their random existence, the most probable random action to take would be to act on themselves, since this is exactly the input that is spatially closer to them. Thus (in contrast to mathematics), self-reference should be a naturally occurring phenomenon, and not the exception to the rule.

\smallskip
 A Darwinian theoretical model for learning in the human brain, has been proposed by various authors \cite{adams, edelman, fernando}.
 
\smallskip
In \cite{watson}, the authors suggest that a gene regulation network has the potential to develop `recall' capabilities normally reserved for cognitive systems.
 
\smallskip Structures that may be the result of sequential or parallel diagonalization are apparent in the human brain:

 The size of the intersection of two assemblies has been shown in experiments to represent the extent to which memories co-occur or concepts are related; the phenomenon is called {\em association of assemblies} \cite{anari}. 

 It has been shown that whenever medial temporal lobe neurons respond to more than one concept, these concepts are typically related. Furthermore, the degree of association between concepts could be successfully predicted based on the neurons response patterns \cite{falco}.

\smallskip
Language is obviously constructed by us in order to communicate. On the other hand, one can easily observe the similarity in structure between language and the encoding of concepts of the real world in the human brain. (See for example \cite{frankland, papadimitriou, ding}). It is therefore very natural to conclude that the creator of both representations is the same. About the possible relation between thinking and language see also \cite{piantadosi2} and the references therein.

\smallskip
(For the definition and related notions of {\em recursive functions} see for example \cite{boolos}). As we have roughly  seen in Example 1 of page \pageref{recursion} and more thoroughly in section \ref{permanentDecisions}, page \pageref{permanentDecisions}, self-editing computations are closed under primitive recursion. It is easy to see that they are also closed under composition and using test algorithms renders them closed under the minimization operator. Thus, according to the theory of recursive functions, a self-editing code requires very few initial functions to be able to compute theoretically any computable function. It is therefore natural, assuming the validity of the theory to human brain, to see simple learning steps to be accomplished by mere strengthening of the synapses, something that could correspond either to composition or to primitive recursion. (However see also \cite{dranovsky}). 

\smallskip
The idea of guessing a differentiating code out of a (finite) set of samples is not new in bibliography: Leslie Valiant \cite{valiant} has proposed a theory for learning, namely
{\em probably approximately correct learning} which is a framework for mathematical analysis of machine learning. The learner receives samples and must select a generalization function (called the {\em hypothesis}) from a certain class of possible functions. The goal is that, with high probability (the ``probably" part), the selected function will have low generalization error (the ``approximately correct" part). 

In \cite{bramley} the authors cover new work that casts human learning as program induction. They argue that the notion that the mind approximates rational (Bayesian) inference is insufficient due to the fact that natural learning contexts are typically much more open-ended--there are often no clear limits on what is possible, and initial proposals often prove inadequate. Recent work has begun to shed light on this problem via the idea that many aspects of learning can be better understood through the mathematics of program induction \cite{chater, lake}. People are demonstrably able to compose hypotheses from parts \cite{goodman, piantadosi2, schulz} and incrementally grow and adapt their models of the world \cite{bramley2}. A number of recent studies has formalized these abilities as program induction, using algorithms that mix stochastic recombination of primitives with memorization and compression to explain data \cite{dechter, ellis, romano}, ask informative questions \cite{rothe}, and support one- and few-shot-inferences \cite{lake}. Program induction is also proving to be an important notion for understanding development and learning through play \cite{sim} and the formation of geometric understanding about the physical world \cite{amalric}. 

\smallskip
\label{hebbianPlasticity} As mentioned earlier, it is easy to see that hebbian plasticity \cite{hebb}, can be thought of as a consequence of diagonalization. Indeed, assuming that two neurons fire together repeatedly,  the permanent decision to strengthen their synapse, should be in accordance with the self-memory.

\smallskip
There are findings \cite{buzsaki2} showing that the brain encodes for the direction of the head, perhaps also of other parts of the body. This seems to be interestingly related to self-editing.

\smallskip
The self-reference effect is a phenomenon studied in psychology \cite{sre1, sre2}, according to which, a person can better recall information in the case where this information has been linked to the self. 

Using a different approach than the one of the present work, Y. Schmidhuber, utilizes the meta-learning capabilities of self-reference in his work. See for example \cite{schmid}.

\smallskip
 The next two arguments are closely related to (and explained better by) the last one:
 
  The well known existence of unconscious parts of ourselves, may be easily explainable by {\em stabilization} about which we have already talked in page \pageref{stabilization}, i.e. the permanent decision of a self-editing code, to keep unaltered parts of its code. Such a decision can result from diagonalization, assuming that there is no successful attempt in the history of a self-editing code to change this part of itself. 

\smallskip
What about emotions? Other than the basic pair of pleasure and dissatisfaction (we 'll talk about them immediately below), one can notice the influence that they represent in a self-editing system. For example, it is easy to observe that {\em fear} can make us function in shorter time cycles, predicting and remembering thus in shorter intervals. This could easily be explained by the fact that under the regime of fear, one should review and change well learned behaviors (which should be established in shorter cycles), in order to adapt them in the new situation.

Let us assume for the moment that `ego' (the greek and latin word for `I') can be defined as the part of ourselves that experiences emotions. Notice that among them {\em pleasure} and {\em dissatisfaction} are the most central. It is evident, both by the way we use education and by our experience, that both these emotions can be used to accomplish learning functionalities.  So, one can conclude, based on the above definition, that `ego' (i.e. the {\em self-referential} part of ourselves) is exactly the part that is intended to accomplish the functions of exploration and learning.

\bigskip
{\bf Ethical statement:} Although it is evident that every research result should be used only for the purpose of a better life for everyone living it, reality most often falsifies this optimistic statement. Consequently, it is a fundamental principle for me, to make clear that I strongly desire, even demand, that every particular concept, result or application of the present work to be used exclusively for the aforementioned purpose. Especially applications that promote greediness, economical or any other kind of violence against life, should be thought as prohibited according to my intentions.

Moreover, I strongly encourage researchers, to state their preferences about the potential applications of their work. I think of this as an obligation of the scientific community to the living world.


\smallskip

\begin{thebibliography}{9}
	
	\bibitem{abeles}
	Abeles M. 
	\textit{Corticonics: Neural Circuits of the Cerebral Cortex.} 1991, Cambridge University Press.
	
	\bibitem{adams}
	Adams P.
	Hebb and Darwin.
	\textit{J. Theoret. Biol.} 1998; {\bf 195}: 419-438
	
	\bibitem{agbehe}
	Aguilar L, Bennati S, Helbing D.
	How learning can change the course of evolution.
	\textit{PLoS ONE} 2019 {\bf 14}(9)
	
	\bibitem{amalric}
		Amalric M, Wang L, Pica P, Figueira S, Sigman M, Dehaene S.
		The language of geometry: Fast comprehension of geometrical primitives and rules in human adults and preschoolers.
	\textit{PLoS Comput. Biol.}, 2017.
	
	\bibitem{anari}
	Anari N, Daskalakis C, Maass W, Papadimitriou C.H, Saberi A, Vempala S.
	Smoothed analysis of discrete tensor decomposition and assemblies of neurons
	in \textit{Advances in Neural Information Processing Systems31: Annual Confeerence on Neural Information Processing Systems 2018, NeurlPS 2018,} 
	Bengio S, Wallach H.M, Eds (Curran Associates, 2018), pp 10880-10890.
	
	\bibitem{bar}
	Bar M. The proactive brain: using analogies and associations to generate predictions. 
	\textit{Trends Cogn Sci.} 2007 {\bf 11}, 280-289.
	
	\bibitem{barton}
	Barton N.H, Novak S, Paix\~ao T.
	Diverse forms of selection in evolution and computer science.
	\textit{Proc. of National Academy of Sciences U.S.A.} 2014;{\bf 111}:10398-10399
	
	\bibitem{bell}
	Bell G. 
	\textit{The Masterpiece of Nature: The Evolution and Genetics of Sexuality.}
	University of California Press, Berkeley, CA, 1982.
	
	\bibitem{boolos}
	Boolos G. S, Burgess J. P, Jeffrey R. C.
	\textit{Computatability and Logic}, 
	Cambridge University Press, 2007.
	
	\bibitem{bramley2}
	Bramley N. R, Dayan P, Griffiths T, Lagnado D.
	Formalizing Neurath's Ship: Approximate Algorithms for Online Causal Learning.
	\textit{Psychological review}, 2017.
	
	
	\bibitem{bramley}
	Bramley N. R, Schulz E, Xu F, Tenenbaum J.
	Learning as program induction.
	\textit{Cognitive Science} 2018.
	
	\bibitem{buzsaki}
	Buzs\'aki G. 
	Neural syntax: Cell assemblies, synapsembles, and readers. 
	\textit{Neuron} 2010 {\bf 68,} 362-385
	
	\bibitem{buzsaki2}
	Buzs\'aki G.
	\textit{The Brain from Inside Out} 2019, Oxford University Press.
	
	\bibitem{calderon-franco}
	Calder\'on-Franco D, van Loosdrecht M.C.M, Abeel T, Weissbrodt D.G.
	Free-floating extracellular DNA: Systematic profiling of mobile genetic elemants and antibiotic resistance from wastewater.
	\textit{Water Research} 2021 {\bf 189} 116592.
	
	
	
	\bibitem{chastain}
	Chastain E, Livnat A, Papadimitriou C, Vazirani U.
	Algorithms, games, and evolution.
	\textit{Proc. of National Academy of Sciences U.S.A.} 2014;{\bf 111}:10620-10623
	
	\bibitem{chater}
	Chater N, Oaksford M.
	Programs as Causal Models: Speculations on Mental Programs and Mental Representation.
	\textit{Cogn. Sci.} 2013.
	
	
	
	\bibitem{clune2}
	Clune J, Mouret J-B, Lipson H. The evolutionary origins of modularity. 
	\textit{Proc R Soc B} 2013 280: 20122863. http://dx.doi.org/10.1098/rspb.2012.2863
	
		
	\bibitem{dasgupta}
	Dasgupta S, Stevens C.F, Navlakha S.
	A neural algorithm for a fundamental computing problem. 
	\textit{Science} 2017 {\bf 358}, 793-796.
	
	\bibitem{dechter}
	Dechter E, Malmaud J, Adams R, Tenenbaum J.
	Bootstrap Learning via Modular Concept Discovery.
	\textit{IJCAI}, 2013.
	
	\bibitem{ding}
	Ding N, Melloni L, Zhang H, Tian X, Poeppel D.
	Cortical Tracking of Hierarchical Linguistic Structures in Connected Speech.
	\textit{Nat. Neurosci.} 2016 {\bf 19}, 158-164.
	
	\bibitem{dranovsky}
	Dranovsky A, Picchini A. M, Moadel T, Sisti A. C, Yamada A, Kimura S, Leonardo E. D, Hen R.
	Experience dictates stem cell fate in the adult hippocampus. 
	\textit{Neuron} 2011 {\bf 70}(5): 908-923.
	
	\bibitem{edelman}
	Edelman G. M. 
	\textit{Neural Darwinism. The Theory of Neuronal Group Selection} 1987, New York: Basic Books.
	
	\bibitem{eichenbaum}
	Eichenbaum H. Barlow versus Hebb: When is it time to abandon the notion of feature detectors and adopt the cell assembly as the unit of congnition? 
	\textit{Neurosci. Lett.} 2018 {\bf 680,} 88-93.
	
	\bibitem{ellis}
	Ellis K, Dechter E, Tenenbaum J.
	Dimensionality Reduction via Program Induction.
	\textit{AAAI Spring Symposia}, 2015.
	
	\bibitem{espinosa-soto}
	Espinosa-Soto C, Wagner A. Specialization can drive the evolution of modularity.
	\textit{PLoS Comput. Biol.} 2010 {\bf 6}, e1000719.
	
		\bibitem{falco}
	Falco E, Ison M.J, Fried I, Quian Quiroga R.
	Long-term coding of personal and universal associations underlying the memory web in the human brain.
	\textit{Nat. Commun.} 2016 {\bf 7,} 13408.
	
	\bibitem{fernando}
	Fernando C, Szathm\'ary E, Husbands P.
	Selectionist and evolutionary approaches to brain function: a critical appraisal
	\textit{Front. Comput. Neurosci.} 2012; {\bf 6}:24
	
	\bibitem{frank}
	Frank S.A. 
	The design of natural and artificial adaptive systems.
	in: Rose M.R. Lauder G.V. 
	\textit{Adaptation.}
	Academic Press, 
	1996: 451-505
	
	\bibitem{frankland}
	Frankland S.M, Greene J.D. 
	An architecture for encoding sentence meaning in left mid-superior temporal cortex.
	2015 \textit{Proc. Natl. Acad. Sci. U.S.A.} {\bf 112} 11732-11737.
	
	\bibitem{fryxell}
	Fryxell K.J, Moon W-J. 
	CpG mutation rates in the human genome are highly dependent on local GC content.
	\textit{Molecular Biology and Evolution.}
    2005, {\bf 22}, 650-658.
	
	\bibitem{facvar}
	Gerhart John, Kirschner Marc. 
	The theory of facilitated variation.
	\textit{Proceedings of the National Academy of Sciences} 2007 {\bf 104}(1):8582-8589.
	

	
	\bibitem{fischer}
	Fisher R.A. 
	\textit{The Genetical Theory of Natural Selection.}
	The Clarendon Press, Oxford, U.K.,1930.
	
	\bibitem{goodman}
	Goodman N.D, Tenenbaum J, Feldman J, Griffiths T.
	A Rational Analysis of Rule-Based Concept Learning.
	\textit{Cogn. Sci.}, 2008.
	
	
	
	\bibitem{graur}
	Graur D, Li W.-H.
	\textit{Fundamentals of Molecular Evolution.}
	Sinauer Associates, Sunderland, MA, 2000.
	
	\bibitem{harris}
	Harris K.D. Neural signatures of cell assembly organization. \textit{Nat. Rev. Neurosci.} 2005 {\bf 6} 399-407
	
	
	
	\bibitem{hauser}
	Hauser M.D, Chomsky N, Fitch W.T.
	The faculty of language: What is it, who has it, and how did it evolve?
	\textit{Science} 2002 {\bf 298}, 1569-1579
	
	\bibitem{hebb}
	Hebb D.O, \textit{The Organization of Behavior: A Neuropsychological Theory} 1949, Wiley, New York, NY.
	
	\bibitem{analogy}
	Hofstadter Douglas. Analogy as the Core of Cognition, in Dedre Gentner, Keith Hoyoak and Boicho Kokinov (eds.) 
	\textit{The Analogical Mind: Perspectives from Cognitive Science}, Cambridge, MA: The MIT Press/Bradford Book, 2001, pp. 499-538.
	
	
	
	\bibitem{kashtan}
	Kashtan N, Alon U. Spontaneous evolution of modularity and network motifs.
	\textit{Proc. Natl. Acad. Sci. USA} 2005 {\bf 102} 13773-13778.
	
	\bibitem{kashtan2}
	Kashtan N, Noor E, Alon U.
	Varying environments can speed up evolution.
	\textit{Proc. Natl Acad. Sci. USA} {\bf 104}, 13711-13716.
	
	\bibitem{kauffman}
	Kauffman S.A.
	Origins of Order: Self-Organization and Selection in Evolution, 1993, Oxford, U.K., Oxford University Press.
	
	\bibitem{kirchner}
	Kirchner M, Gerhart J.
	Evolvability.
	\textit{Proceedings of National Academy of Sciences U.S.A.} 1998; {\bf 95}: 8420-8427
	
	\bibitem{kleene}
	Kleene Stephen Cole.
	\textit{Introduction to Metamathematics}.
	Wolters-Noordhoff publishing - Groningen North-Holland publishing company - Amsterdam New York, 1952.
	
	\bibitem{sre1}
	Klein S.B, Kihlstrom J.F. 
	Elaboration, organization, and the self-reference effect in memory. 
	\textit{J Exp Psychol Gen.} 
	1986 Mar;{\bf115}(1):26-38. doi: 10.1037//0096-3445.115.1.26. PMID: 2937872.
	
	
	\bibitem{kouvaris}
	Kouvaris K, Clune J, Kounios L, Brede M, Watson R. A.
	How evolution learns to generalise: Using the principles of learning theory to understand the evolution of developmental organisation.
	\textit{PLoS Comput Biol} 2017 {\bf 13}(4)
	
	\bibitem{lake}
	Lake B, Salakhutdinov R, Tenenbaum J.
	Human-level concept learning through probabilistic program induction.
	\textit{Science}, 2015.
	
	\bibitem{laland}
	Laland K.N, Uller T, Feldman M.W, Sterelny K, M\"uller G.B, Moczed A, Jablonka E, Odling-Smee J.
	The extended evolutionary synthesis: its structure assumptions and predictions. 
	\textit{Proc. R. Soc. B} 2015 {\bf 282}: 20151019. http://dx.doi.org/10.1098/rspb.2015.1019
	
	\bibitem{levin}
	Levin S.A.
	Ecosystems and the biosphere as complex adaptive systems.
	\textit{Ecosystems.} 1998; {\bf 1}: 431-436
	
	\bibitem{lewontin}
	Lewontin R.C, Hubby J.L. A molecular approach to the study of genic heterozygosity in natural poputations; amount of variation and degree of heterozygosity in natural populations of {\em Drosophila pseudoobscura.} \textit{Genetics} {\bf 54}, 1966, pp 595--609.
	
	\bibitem{livnat2}
	Livnat A. Interaction-based evolution: How natural selection and nonrandom mutation work together.
	\textit{Biology Direct} 8, 1 (2013), 24.
	
	\bibitem{livnat}
	Livnat A, Papadimitriou C.
		Sex as an Algorithm: The Theory of Evolution Under the Lens of Computation.
	\textit{Communication of the ACM} 2016 {\bf 59}(11) 84-93.
	
	\bibitem{margulis}
	Margulis L.
	Origins of species: acquired genomes and individuality.
	\textit{BioSystems.} 1993; {\bf 31}; 121-125
	
	\bibitem{miller}
	Miller J.E. K., Ayzenshtat I, Carrillo-Reid L, Yuste R.
	Visual stimuli recruit intrinsically generated cortical ensembles.
	\textit{Proc. Natl. Acad. Sci. U.S.A.} 2014 {\bf 111}, E4053-E4061.
	
	\bibitem{moschovakis}
	Moschovakis Y.N.
	\textit{Notes on Set Theory}.
	Springer, 2006.
	
	
	\bibitem{neisser}
	Neisser U.
	\textit{Cognition and reality: principles and implications of cognitive psychology.} San Fracisco: Freeman, 1976.
	
	\bibitem{nevo}
	Nevo E, Beiles A, Ben-Shlomo R. The evolutionary significance of genetic diversity: Ecological, demographic and life history correlates.
	\textit{Lecture Notes in Biomathematics} {\bf 53}, 1984.
	
	\bibitem{papadimitriou2}
	Papadimitriou C, Steiglitz K. 
	\textit{Combinatiorial Optimization: Algorithms and Complexity.} Dover, 1998.
	
	\bibitem{papadimitriou}
	Papadimitriou C.H, Vempala S. S, Mitropolsky D, Collins M, Maass W.
	Brain computation by assemblies of neurons.
	\textit{Proceedings of the National Academy of Sciences} 2020, {\bf 117}(25) pp 14464-14472; doi:10.1073/pnas.2001893117
	
	\bibitem{pavlicev}
	Pavlicev M, Cheverud J. M, Wagner G. P,
	Evolution of adaptive phenotypic variation patterns by direct selection for evolvability
	\textit{Proc. R. Soc. B Biol. Sci.} 2011;{\bf 278}: 1903-1912
	
		
	\bibitem{piantadosi2}
	Piantadosi S. T, Tenenbaum J. B, Goodman N. D. The logical primitives of thought: Empirical foundations for compositional cognitive models. \textit{Psychol. Rev.} 2016 {\bf 123,} 392-424
	
	
	\bibitem{pigliucci}
	Pigliucci M.
	Is evolvability evolvable?.
	\textit{Nat. Rev. Genet.} 2008; {\bf 9}: pp75-82
	
	\bibitem{ranhel}
	Ranhel J. Neural Assembly Computing.
	\textit{IEEE Trans. Neural Netw. Learn. Syst.} 2012 {\bf 23}, 916-927.
	
	
	
	\bibitem{riedl}
	Riedl R.J.
	A systems-analytical approach to macroevolutionary phenomena.
	\textit{Q. Rev. Biol.} 1977; {\bf 52}: 351-370.
	
	\bibitem{romano}
	Romano S, Salles A, Amalric M, Dehaene S, Sigman M, Figueira S.
	Bayesian selection of grammar productions for the language of thought.
	bioRxiv, 2017.
	
	\bibitem{rothe}
	Rothe A, Lake B, Gureckis T.
	Question Asking as Program Generation.
	\textit{NIPS}, 2017.
	
	\bibitem{schmid}
	Schmidhuber J. G\"odel Machines: Fully Self-referential Optimal Universal Self-improvers. In: {\em Artificial General Intelligence. Congnitive Technogies.} Editors: Goertzel B., Pennachin C. Springer, Berlin, Heidelberg, 2007.
	
	\bibitem{schulz}
	Schulz E, Tenenbaum J, Duvenaud D, Speekenbrink M, Gershman S.
	Compositional Inductive Biases in Function Learning.
	biorxiv, 2016.
	
	\bibitem{sim}
	Sim A, Xu F.
	Learning Higher-Order Generalizations Through Free Play: Evidence From 2- and 3-Year-Old Children.
	\textit{Developmental psychology}, 2017.
	
	\bibitem{smullyan}
	Smullyan R.M. \textit{Diagonalization and Self-Reference.} Oxford science publications, 1994.
	
	
	
	\bibitem{stearns}
	Stearns S.C, Hoekstra R.F.
	\textit{Evolution: An Introduction.}
	Oxford University Press, New York, 2005.
	
	\bibitem{szathmary}
	Szathm\'ary E.
	Toward major evolutionary transitions theory 2.0.
	\textit{Proc. Natl. Acad. Sci. U.S.A.} 2015; {\bf 112}:10104-10111
	
	
	\bibitem{sre2}
	Symons C.S, Johnson B.T. 
	The self-reference effect in memory: a meta-analysis. 
	\textit{Psychol Bull.} 
	1997 May; {\bf 121}(3):371-94. doi: 10.1037/0033-2909.121.3.371. PMID: 9136641.
	
	\bibitem{valiant}
	Valiant L. 
	\textit{Probably Approximately Correct: Nature's Algorithms for Learning and Prospering in a Complex World.} Basic Books, 2013.
	
	
	
	
	
	\bibitem{wagner-altenberg}
	Wagner, G.P, Altenberg, L.
	Complex adaptation and the evolution of evolvability.
	\textit{Evolution} 1996 {\bf 50}, pp. 967-976.
	
	\bibitem{wagner2}
	Wagner G, Mezey J, Calabretta R.
	\textit{Modularity: understanding the development and evolution of complex natural systems.} Natural selection and the origin of modules. 2001 Cambridge, MA: MIT Press.
	
	\bibitem{wagner}
	Wagner G.P, Pavlicev M, Cheverud J.M.
	The road to modularity.
	\textit{Nat. Rev. Genet.} 2007 {\bf 8}, 921-931.
	
	\bibitem{watson}
	Watson R.A, Buckley C.L, Mills R, Davies A.
	Associative memory in gene regulation networks.
	in: Fellermann H. \textit{Proceedings of the Artificial Life Conference XII.} MIT Press, 2010: 194-202
	
	\bibitem{intev}
	Watson A.R, Szathm\'ary E. How Can Evolution Learn? 
	\textit{Trends in Ecology \& Evolution,} 2015.
	
	\bibitem{west-eberhard}
	West-Eberhard M.J. 
	Phenotypic Plasticity and the Origins of Diversity.
	\textit{Annual Review of Ecology and Systematics,} {\bf 20} (1989), pp. 249-278.
	
	\bibitem{wright}
	Wright S.
	The distribution of gene frequencies in populations.
	\textit{Proceedings of the National Academy of Sciences U.S.A.} 1937 {\bf 23}(6), 307320.
	
	
	
\end{thebibliography}
  \end{document}